\documentclass[12pt]{article}
\usepackage[centertags]{amsmath}
\usepackage{amsfonts}
\usepackage{amssymb}
\usepackage{amsthm}
\usepackage{cite}
\usepackage{graphicx}
\usepackage{etoolbox}
\usepackage{slashed}
\usepackage[vcentermath]{youngtab}
\usepackage{multirow}
\usepackage{array}
\usepackage{tabularx}

\newcommand{\beq}{\begin{equation}}
\newcommand{\eeq}[1]{\label{#1}\end{equation}}
\newcommand{\bea}{\begin{eqnarray}}
\newcommand{\eea}[1]{\label{#1}\end{eqnarray}}

\patchcmd{\subequations}{}%
{}{}{}

\def\a{\alpha}
\def\b{\beta}
\def\g{\gamma}

\def\d{\delta}
\def\D{\Delta}
\def\e{\epsilon}
\def\ve{\varepsilon}

\def\h{\eta}

\def\l{\lambda}
\def\L{\Lambda}
\def\m{\mu}
\def\n{\nu}

\def\r{\rho}

\def\s{\sigma}

\def\U{\Upsilon}

\def\F{\Phi}

\def\Ps{\Psi}

\def\mc{\mathfrak{c}}
\def\mD{\mathcal{D}}
\def\mm{\mathfrak{m}}
\def\mg{\mathfrak{g}}
\def\mf{\mathfrak{f}}

\def\de{\partial}
\def\nb{\nabla}
\def\bbox{\de^2}
\def\div{d\!\cdot\!\partial}
\def\grad{u\!\cdot\!\partial}
\def\spin{u\!\cdot\!d}

\def\dbbox{\nb^2}
\def\ddiv{d\!\cdot\!\nabla}
\def\dgrad{u\!\cdot\!\nabla}
\def\des{\slashed{\partial}}
\def\ds{\slashed{d}}
\def\us{\slashed{u}}
\def\nbs{\slashed{\nabla}}
\def\dal{\underline{g}_0}
\def\dalh{\underline{\hat{g}}_0}
\def\dalb{\underline{\bar{g}}_0}
\def\doub{\left[
            \begin{array}{c}
              \Phi \\
              \lambda \\
            \end{array}
          \right]}

\def\dir{\underline{f}_0}
\def\dirh{\underline{\hat{f}}_0}

\def\douf{\left[
            \begin{array}{c}
              \Psi \\
              \ve \\
            \end{array}
          \right]}
\def\Fpm{\F^{(k+1)}_{s}}
\def\lpm{\l_{s-k-1}}
\def\Pspm{\Ps^{(k+1)}_{n}}
\def\vepm{\ve_{n-k-1}}
\def\doupmb{\left[
            \begin{array}{c}
              \F_s^{(k+1)} \\
              \l_{s-k-1} \\
            \end{array}
          \right]}
\def\doupmf{\left[
            \begin{array}{c}
              \Ps_n^{(k+1)} \\
              \ve_{n-k-1} \\
            \end{array}
          \right]}

\def\Ddiv{d\!\cdot\!\D}
\def\Dgrad{u\!\cdot\!\D}
\def\Ds{\slashed{\D}}

\def\oone{\mathcal{O}\!\left(\tfrac{1}{\L}\right)}
\def\otwo{\mathcal{O}\!\left(\tfrac{1}{\L^2}\right)}

\def\otwob{\mathcal{O}\!\left(\tfrac{1}{\bar{\L}^2}\right)}
\def\ooneq{\mathcal{O}\!\left(q\right)}
\def\otwoq{\mathcal{O}\!\left(q^2\right)}
\def\rad{\tfrac{1}{L^2}}
\def\radf{\tfrac{1}{L}}

\begin{document}
\numberwithin{equation}{section}
\setlength{\topmargin}{-1cm} \setlength{\oddsidemargin}{0cm}
\setlength{\evensidemargin}{0cm}

\begin{titlepage}
\begin{center}
{\Large \bf The Involutive System of Higher-Spin Equations}

\vspace{20pt}

{\large Rakibur Rahman$^{\,a,b}$}

\vspace{12pt}
$^a$ Max-Planck-Institut f\"ur Gravitationsphysik (Albert-Einstein-Institut)\\
     Am M\"uhlenberg 1, D-14476 Potsdam-Golm, Germany\\
\vspace{6pt}
$^b$ Department of Physics, University of Dhaka, Dhaka 1000, Bangladesh\\
\end{center}
\vspace{20pt}

\begin{abstract}

We revisit the problem of consistent free propagation of higher-spin fields in nontrivial
backgrounds, focusing on symmetric tensor(-spinor)s. The Fierz-Pauli equations for massive
fields in flat space form an involutive system, whose algebraic consistency owes to certain
gauge identities. The zero mass limit of the former leads directly to massless higher-spin
equations in the transverse-traceless gauge, where both the field and the gauge parameter
have their respective involutive systems and gauge identities. In nontrivial backgrounds,
it is the preservation of these gauge identities and symmetries that ensures the correct
number of propagating degrees of freedom. With this approach we find consistent sets of
equations for massive and massless higher-spin bosons and fermions in certain
gravitational/electromagnetic backgrounds. We also present the involutive system of partially
massless fields, and give an explicit form of their gauge transformations. We consider the Lie
superalgebra of the operators on symmetric tensor(-spinor)s in flat space, and show that in AdS
space the algebra closes nonlinearly and requires a central extension.

\end{abstract}

\end{titlepage}

{\footnotesize\tableofcontents}

\newpage
\section{Introduction}\label{sec:intro}

The construction of consistent interacting theories of higher-spin fields is a difficult task.
Generic interactions of massless fields are incompatible with gauge invariance, and this fact
gives rise to various no-go theorems~\cite{Old,gpv,Aragone,ww,New}. For massive fields, when
interactions are turned on, the dynamical equations and constraints may either lose algebraic
consistency~\cite{FP} or start propagating unphysical/superluminal modes~\cite{vz,sham,kob,d3}.
These pathologies show up even for a much simpler setup that we would like to consider in this
article: the free propagation of higher-spin fields in nontrivial backgrounds (see~\cite{Rahman:2015pzl}
for a recent review).

In this article, we employ the metric-like formulation, where the degrees of freedom (DoF) of higher-spin
particles are encoded in symmetric tensors and tensor-spinors. The flat-space free Lagrangians
and the equations of motion (EoM) are well known for massive and massless metric-like
fields~\cite{Rahman:2015pzl}. In nontrivial backgrounds, however, consistent propagation is not at all
automatic; one must ensure among other things that only the physical modes propagate and that their
propagation remains causal. This is the weakest link of the Lagrangian formulation, for both
massive~\cite{vz,sham,kob,d3} and massless fields~\cite{Henneaux:2013vca}, since the problems become
manifest only at the EoM level. Moreover, the EoM's often turn out to be surprisingly simple, but this
simplicity is obscured at the Lagrangian level~\cite{AN,Klishevich,PRS,PR3half}.

It is therefore desirable to study the propagation of higher-spin fields
solely at the EoM level, without recourse to the Lagrangian formulation. This is where
the involutive properties of higher-spin equations come into play (see Appendix~\ref{sec:appendix}
for an exposition of involutive systems). Devoid of a parent Lagrangian, the mutual
compatibility of the dynamical equations and constraints/gauge-fixing conditions in a nontrivial
background is no longer guaranteed. The good news is that this can be duly taken care of by the ``gauge
identities'' of the involutive system. In fact, in the involutive approach, all the consistency
issues are under full control, so that one may systematically deform the flat-space system of
higher-spin equations. This ``involutive deformation method'' has already been employed for the free
propagation of massive bosons in various backgrounds~\cite{Cortese:2013lda,Kulaxizi:2014yxa,Cortese:2017ieu}.
In this article, we would like to extend this approach to fermions as well as to gauge fields.

The organization of this article is as follows. The remaining of this section gives a brief account
of the operator formalism$-$a handy computational tool to be used throughout the article. Section~\ref{sec:mb}
deals with the Fierz-Paui system for massive bosons, and rederives its involutive deformations in gravitational
and electromagnetic backgrounds using the elegant operator formalism. The extension of this construction to
massive fermions is presented in Section~\ref{sec:mf}. Sections~\ref{sec:gb} and~\ref{sec:gf} respectively
consider gauge bosons and fermions, where we first present the flat-space involutive systems in the
transverse-traceless gauge, obtained in the zero mass limits of their massive counterparts. Then we construct
their respective deformations in gravitational and electromagnetic backgrounds$-$a task made challenging
by ``unfree'' gauge symmetries~\cite{unfree}, whose parameters themselves are governed by involutive
systems. Section~\ref{sec:PMAdS} analyzes the involutive systems of partially massless bosons and fermions
along with their gauge transformations. In Section~\ref{sec:Consistency}, we show how the various operators
acting on symmetric tensor(-spinor)s  in AdS space form a nonlinear Lie superalgebra with a central charge.
Some concluding remarks are made in Section~\ref{sec:remarks}, in particular about the possible r\^ole of mixed-symmetry
fields. Three appendices provide brief accounts of involutive systems and deformations,
and some technical details.

\subsubsection*{The Operator Formalism}

The operator formalism introduces auxiliary tangent-space variables $u^a$ and their derivatives:
$d_a\equiv\frac{\de}{\de u^a}$, where fiber (world) indices are denoted by lower case Roman (Greek) letters.
The vielbein $e^{\m}_a(x)$ and its inverse $e_{\m}^{a}(x)$ give the contracted auxiliary variables:
\beq u^\m\equiv e^{\m}_{a}(x)u^{a},\qquad d_{\m}\equiv e_{\m}^{a}(x)d_a,\eeq{1.1}
which comprise a set of oscillators that satisfies the Heisenberg algebra:
\beq [u^\m,u^\n]=0,\qquad [d_\m,d_\n]=0,\qquad [d_\m,u^\n]=\d_\m^\n.\eeq{1.2}

A symmetric rank-$s$ tensor $\F_{\m_1\cdots\m_s}(x)$ denotes a spin-$s$ bosonic field, while a symmetric rank-$n$
tensor-spinor $\Ps_{\m_1\cdots\m_n}(x)$, with the spinor index kept implicit, denotes a fermionic field of spin
$s=n+\tfrac{1}{2}$. They are represented respectively by the generating functions:
\beq \F(x,u)=\tfrac{1}{s!}\,\F_{\m_1\cdots\m_s}(x)\,u^{\m_1}\,\cdots\,u^{\m_s},\qquad\Ps(x,u)=\tfrac{1}{n!}\,
\Ps_{\m_1\cdots\m_n}(x)\,u^{\m_1}\,\cdots\,u^{\m_n},\eeq{1.3}
The commutator of covariant derivatives acts on them in the following way:
\bea &[\nb_\m,\nb_\n]\,\F=R_{\m\n\r\s}(x)u^\r d^\s\F,&\label{1.4}\\
&[\nb_\m,\nb_\n]\,\Ps=R_{\m\n\r\s}(x)u^\r d^\s\Ps+\tfrac{1}{4}R_{\m\n\r\s}(x)\g^\r\g^\s\Ps,&\eea{1.5}
with $\g^\m\equiv e^\m_a(x)\g^a$, where $\g^a$ are the tangent-space gamma matrices. It is important to note that
the vielbein postulate results in the following vanishing commutators:
\beq [\nb_\m,u^\n]=0,\qquad [\nb_\m,d_\n]=0,\qquad [\nb_\m,\g^\n]=0.\eeq{1.6}

The index operator is: $N\equiv\spin=u^\m d_\m$, where a ``dot'' stands for the contraction
of a pair of indices. For any operator $\hat O$, there is a corresponding weight $w$ of $N$, given by:
\beq [N,\hat O]=w\hat O.\eeq{1.8}
The weight $w$ is an intrinsic property, which counts the tensor rank of the operator.

The case of flat space is special, where the vielbein $\hat{e}^{\,a}_\m$ satisfies:
$\hat{e}^{\,a}_\m\hat{e}_{\n a}=\h_{\m\n}$. Then, it suffices to consider only world indices that can be
lowered and raised by the Minkowski metric and its inverse. In the absence of any gauge connections, one
is left only with partial derivatives $\de_\m$ that are of commuting nature: $[\de_\m,\de_\n]=0$.

\newpage
\section{Massive Bosonic Fields}\label{sec:mb}

In this section, we study the Fierz-Paui system for totally-symmetric massive bosons in the operator
formalism. We start with the free propagation in Minkowski background, where we properly identify all
the gauge identities of the involutive system. Then, the involutive deformations in
gravitational/electromagnetic backgrounds~\cite{Cortese:2013lda,Kulaxizi:2014yxa,Cortese:2017ieu} are
rederived, rather more elegantly, using the operator formalism. Despite having no new results, this
section will be immensely useful for the sake of familiarity with the concepts and methodology.

\subsection{Minkowski Background}\label{sec:mbflat}

The Fierz-Pauli conditions for a symmetric bosonic field of mass $m$ in flat space
involve the Klein-Gordon, divergence and trace operators~\cite{Rahman:2015pzl}, comprising the set:
\beq \mathcal{G}=\left\{g_0,g_1,g_2\right\},\eeq{2.1}
where a subscript gives the negative weight ($-w$) corresponding to an operator. Table~1 summarizes
the various properties of these operators.
\begin{table}[ht]
\caption{Operators in Bosonic Fierz-Pauli System}
\vspace{6pt}
\centering
\begin{tabular}{c c c c c}
\hline\hline
Operator~~~&~~~Symbol~~~&~~~Definition~~&~~Weight $(w)$~~&~~Derivative Order $(k)$\\ [0.5ex]
\hline
Klein-Gordon & $g_0$ & $\de^2-m^2$ & $~~0$ & 2\\
Divergence & $g_1$ & $\div$ & $-1$ & 1\\
Trace & $g_2$ & $d^2$ & $-2$ & 0\\
\hline\hline
\end{tabular}
\end{table}
\vspace{6pt}

\noindent
Let us now consider the commutators between two \emph{different} operators:
\beq c_1\equiv[g_0,g_1],\qquad c_2\equiv[g_2,g_0],\qquad c_3\equiv[g_1,g_2],\eeq{2.3}
all of which vanish on account of the commutativity of partial derivatives. Moreover,
these linear operators have associative property, so that the Jacobi identity holds:
\beq [g_0,c_3]+[g_1,c_2]+[g_2,c_1]=0.\eeq{2.4}

The Fierz-Pauli equations constitute an \emph{involutive system} of differential
equations~\cite{Inv-book}:
\beq g_0\F=0,\qquad g_1\F=0,\qquad g_2\F=0.\eeq{2.5}
From the point of view of an involutive system, the algebraic consistency of the system~(\ref{2.5})
is taken care of by the \emph{gauge identities}~\cite{KaLySh3} (see also Appendix~\ref{sec:appendix}):
\beq c_1\F=0,\qquad c_2\F=0,\qquad c_3\F=0,\eeq{2.6}
which hold good because $c_i$'s themselves vanish. For the involutive system~(\ref{2.5}), however,
the gauge identities~(\ref{2.6}) are \emph{not irreducible}. To see this, let us define the operator:
\beq j_3\equiv g_2c_1+g_1c_2+g_0c_3.\eeq{2.65}
Then, the Jacobi identity~(\ref{2.4}) implies the following \emph{on-shell} identity:
\beq j_3\F=0.\eeq{2.7}
In other words, given the system of equations~(\ref{2.5}), we have a gauge identity at reducibility order 1.
This exhausts the list of all possible gauge identities for our system.

The system~(\ref{2.5})--(\ref{2.7}) of involutive equations plus gauge identities is of the
kind considered in Appendix~\ref{sec:nogauge}. To check its absolute compatibility and find the
DoF count, let us first give the number of equations at order $k$ in \emph{space-time derivatives}.
For a symmetric boson of rank/spin $s$, in $D$ is space-time dimensions, it is given by:
\beq t_k=\d_k^2\,{D+s-1\choose s}+\d_k^1\,{D+s-2\choose s-1}+\d_k^0\,{D+s-3\choose s-2},\eeq{2.8}
where a weight-$w$ operator acting on a rank-$s$ tensor gives ${D+s+w-1\choose s+w}$ number
of equations. On the other hand, the number of $\mathcal{O}(k)$ gauge identities at reducibility
order $j$ is:
\beq l_{k,\,j}=\d_k^3\d_j^0\,{D+s-2\choose s-1}+\d_k^2\d_j^0\,{D+s-3\choose s-2}
+\left(\d_k^1\d_j^0+\d_k^3\d_j^1\right){D+s-4\choose s-3}. \eeq{2.9}
With the total number of field variables $f={D+s-1\choose s}$, one finds from Eq.~(\ref{A17})
that $\mc=0$, i.e., the bosonic Fierz-Pauli system is absolutely compatible. The physical DoF count
per space-time point, computed from Eq.~(\ref{A18}), turns out to be:
\beq \mathfrak{D}_b(s)=2\,{D-4+s\choose s-1}+{D-4+s\choose s},\eeq{2.12}
which is indeed the correct number of propagating DoF's of a massive spin-$s$ boson~\cite{Rahman:2015pzl}.

\subsection{Gravitational Background}\label{sec:mbgrav}

In order to describe the free propagation of a massive boson in a gravitational background, we would like
to apply the involutive deformation method to the flat-space system of the previous section. As outlined
in Appendix~\ref{sec:invdeform}, the zeroth-order deformations are obtained by replacing ordinary
derivatives by covariant ones: $\de_\m\rightarrow\nb_\m$, while the first-order ones should be linear in
the background curvature tensor, and so on. The most generic deformations of the operators~(\ref{2.1})
take the following form:
\begin{equation}\label{2.13} \begin{split}
\text{Klein-Gordon}:~\hat{g}_0&=\dbbox\!-\!M^2\!+\a_1R_{\m\n\r\s}u^\m u^\r d^\n d^\s
+\a_2R_{\m\n}u^\m d^\n+\a_3 R+\otwo,\\
\text{Divergence}:~\hat{g}_1&=\ddiv+\otwo,\\
\text{Trace}:~\hat{g}_2&=d^2+\otwo,\end{split}\end{equation}
where the deformed mass $M^2$ and the dimensionless operators $\a_1, \a_2, \a_3$ have weight $w=0$, and
the mass scale $\L$ is larger than other scales in the theory. Note that the book-keeping parameter
(see Appendix~\ref{sec:invdeform}) indicating the deformation order is implicit here. The
deformations~(\ref{2.13}), of course, preserve the respective weights $w$ of the operators. Because the
deformations are smooth, $M^2\rightarrow m^2$ in the limit of zero curvature.

Now, we would like to calculate the commutators between two different operators. The technical steps of
the explicit computations of the desired commutators: $[\hat{g}_0,\hat{g}_1]$, $[\hat{g}_1,\hat{g}_2]$
and $[\hat{g}_2,\hat{g}_0]$ are relegated to Appendix~\ref{sec:tech1}. In order for having some deformed
gauge identities in the first place, we should ensure that these commutators close within the given set
of operators. Among other things, we have the following expression:
\beq [\hat{g}_1,\hat{g}_0]=2(\a_1-1)W_{\m\n\r\s}\nb^\m u^\r d^\n d^\s
+\left[(\a_2+1)-\left(\tfrac{2N}{D-2}\right)(\a_1-1)\right]S_{\m\n}\nb^\m d^\n+\cdots,\eeq{master}
where $W_{\m\n\r\s}$ and $S_{\m\n}\equiv R_{\m\n}-\tfrac{1}{D}g_{\m\n}R$ are respectively the Weyl tensor
and the traceless part of the Ricci tensor, and the ellipses stand for other kinds of terms whose explicit
forms do not matter at his point. In particular, some of the latter terms involve the gradient of the
Riemann tensor, which can be decomposed into irreducible Lorentz tensors:
\beq \underbrace{\yng(1)~\otimes~\yng(2,2)}_{\text{Gradient of Riemann}}~=\quad\underbrace{\yng(3,2)}_{X}~~\oplus
~~\underbrace{\yng(3)}_{Y}~~\oplus~~\underbrace{\yng(2,1)}_{Z}~~\oplus~~\underbrace{\yng(1)}_{U}~,\eeq{tableaux}
where, with the convention that (anti)symmetrization of indices has unit normalization,
\begin{equation}\label{XYZUdefined} \begin{split} X_{\m\n\r}{}^{\a\b}~&\equiv~\nb_{(\m}
W_\n{}^\a{}_{\r)}{}^\b-\left(\tfrac{2}{D+2}\right)g_{(\m\n}\nb^\s W_{\r)}{}^{(\a}{}_\s{}^{\b)},\\
Y_{\m\n\r}~&\equiv~\nb_{(\m}R_{\n\r)}-\left(\tfrac{2}{D+2}\right)g_{(\m\n}\nb_{\r)}R,\\
Z_{\m\n\r}~&\equiv~2\nb_{[\r}R_{\n]\m}+\left(\tfrac{1}{D-1}\right)g_{\m[\r}\nb_{\n]}R
+\left(\m\leftrightarrow\n\right),\\U_\m~&\equiv~\nb_\m R.\end{split}\end{equation}
For an arbitrary-spin field in $D\geq4$, it is clear from Eq.~(\ref{master}) that the two
terms on the right hand side must vanish for a gauge identity to hold good; this demands:
\beq \a_1=1,\qquad \a_2=-1.\eeq{2.140}
Then, the explicit form of Eq.~(\ref{master}) reduces to:
\begin{equation}\begin{split}
[\hat{g}_1,\hat{g}_0]&=[M^2,\ddiv]-R\,[\a_3,\ddiv]+\left[\a_3-\tfrac{2(N-1)(N+D-2)}{(D-1)(D+2)}\right]
d\!\cdot\!U+\tfrac{2\left(u^2d\cdot U+u\cdot U\right)}{(D-1)(D+2)}d^2\\&+X_{\m\n\r}{}^{\a\b}
u_\a u_\b d^\m d^\n d^\r+Z_{\m\n\r}\left[\tfrac{2}{3}u^\r d^\m d^\n+\tfrac{1}{3(D-2)}u^\r N d^\m d^\n
+\tfrac{4D-7}{3(D^2-4)}u^\m u^\n d^\r d^2\right]\\&-Y_{\m\n\r}\left[u^\m d^\n d^\r-\tfrac{1}{D-2}
\left(u^\m u^\n d^\r d^2+u^2d^\m d^\n d^\r- 2 u^\m N d^\n d^\r\right)\right]+\otwo.\nonumber
\end{split}\end{equation}
The last two lines in the above equation impose the following constraints:
\beq X_{\m\n\r}{}^{\a\b}=0,\qquad Y_{\m\n\r}=0,\qquad Z_{\m\n\r}=0.\eeq{2.14}
Without constraining the gravitational background any further, we can also choose:
\beq \a_3=\tfrac{2(N-1)(N+D-2)}{(D-1)(D+2)}\,.\eeq{2.15}
Finally, in order to deal with the commutator $[M^2,\ddiv]$, let us assume that the deformed mass $M^2$
is a quadratic polynomial in the index operator $N$:
\beq M^2=m^2+\m^2\left(N^2+\b N+\g\right),\eeq{2.16}
where $\b$ and $\g$ are some numerical constants, and $\m$ is some constant mass parameter that
vanishes in the limit of zero curvature. The justification of such an assumption can only be given
a posteriori, when we consider the massless case. Then, we have:
\beq [M^2,\ddiv]=P(N)\ddiv,\qquad P(N)\equiv-\m^2\left(2N+\b+1\right).\eeq{2.17}
With the choices and constraints~(\ref{2.140})--(\ref{2.17}), the commutator~(\ref{master}) reduces to:
\beq [\hat{g}_1,\hat{g}_0]=\tfrac{2}{(D-1)(D+2)}\left[R\left(2N+D-2\right)\hat{g}_1+\left(u^2d\!\cdot\!U
+u\!\cdot\!U\right)\hat{g}_2\right]+P(N)\hat{g}_1+\otwo.\eeq{2.18}
Similarly, in view of the choices~(\ref{2.140}) and~(\ref{2.15})--(\ref{2.16}), we have the following result:
\beq [\hat{g}_2,\hat{g}_0]=\tfrac{4}{(D-1)(D+2)}\,R\left(2N+D-1\right)\hat{g}_2+Q(N)\hat{g}_2
+\otwo,\eeq{2.19}
where $Q(N)\equiv-2\m^2\left(2N+\b+2\right)$. The third and last commutator takes the simple form:
\beq [\hat{g}_1,\hat{g}_2]=\otwo.\eeq{2.20}

Given the relations~(\ref{2.18})--(\ref{2.20}), we now identify the deformed counterparts of the
commutators appearing in Eq.~(\ref{2.3}). They are:
\bea &\hat{c}_1\equiv[\hat{g}_0,\hat{g}_1]+\tfrac{2}{(D-1)(D+2)}
\left[R\left(2N+D-2\right)\hat{g}_1+\left(u^2d\!\cdot\!U+u\!\cdot\!U\right)\hat{g}_2\right]
+P(N)\hat{g}_1,&\nonumber\\&\hat{c}_2\equiv[\hat{g}_2,\hat{g}_0]-\tfrac{4}{(D-1)(D+2)}\,
R\left(2N+D-1\right)\hat{g}_2-Q(N)\hat{g}_2,&\label{2.21}\\
&\hat{c}_3\equiv[\hat{g}_1,\hat{g}_2].&\nonumber\eea{xxyyzz}
Finally, we identify the deformed version of the operator $j_3$ of Eq.~(\ref{2.65}) with:
\beq \hat{j}_3\equiv\hat{g}_2\hat{c}_1+\hat{g}_1\hat{c}_2+\hat{g}_0\hat{c}_3.\eeq{2.22}
On account of the Jacobi identity among the deformed operators $\left\{\hat{g}_0,\hat{g}_1,\hat{g}_2\right\}$,
we can use the definitions~(\ref{2.21}) to express $\hat{j}_3$ in the following form:
\beq \hat{j}_3=\hat{O}_3\,\hat{g}_0+\hat{O}_2\,\hat{g}_1+\hat{O}_1\,\hat{g}_2,\eeq{2.23}
where $\hat{O}_i$ is an operator of weight $-i$, whose explicit expression is given in Eq.~(\ref{C8}).

Now we are ready to present our deformed involutive system with all the gauge identities. Of course, the system of
equations is given by:
\beq \hat{g}_0\F=0,\qquad \hat{g}_1\F=0,\qquad \hat{g}_2\F=0.\eeq{2.24}
The gauge identities at reducibility order $0$ can be written in the following form:
\beq \hat{c}_1\F=0,\qquad \hat{c}_2\F=0,\qquad \hat{c}_3\F=0,\eeq{2.25}
provided that the $\hat{c}_i$'s, given by Eqs.~(\ref{2.21}), vanish identically. This happens when the $\otwo$
terms in the commutators~(\ref{2.18})--(\ref{2.20}) are zero. Without explicit knowledge of similar
terms in the operators $\left\{\hat{g}_0,\hat{g}_1,\hat{g}_2\right\}$, the latter condition can be ensured by
taking\footnote{Alternatively, when $\otwo$ terms are judiciously included in the equations~(\ref{2.24}),
similar contributions should be absent in the commutators~(\ref{2.18})--(\ref{2.20}) modulo additional on-shell
vanishing terms. This may pose additional constraints on the gravitational background. We would not consider this
possibility.}:
\beq \L\rightarrow\infty.\eeq{2.26}
On account of the relation~(\ref{2.23}), we also have the following on-shell identity:
\beq \hat{j}_3\F=0,\eeq{2.27}
which is the desired gauge identity at reducibility order 1. This completes our involutive deformation analysis
of a free massive boson in a gravitational background.

To summarize, the consistent dynamical equation for a free massive boson reads:
\beq \left(\dbbox\!-\!M^2\!+R_{\m\n\r\s}u^\m u^\r d^\n d^\s-R_{\m\n}u^\m d^\n+\tfrac{2(N-1)(N+D-2)}{(D-1)(D+2)}\,R\right)
\F=0,\eeq{2.28}
where the deformed mass is of the type~(\ref{2.16}). The constraint equations are given by:
\beq \ddiv\F=0,\qquad d^2\F=0.\eeq{2.29}
The involutive nature of this system hinges upon the constraints~(\ref{2.14}) on the background.
This result essentially captures those already found in~\cite{Cortese:2013lda,Kulaxizi:2014yxa},
and is valid for arbitrary spin in $D\geq4$. Below we consider some important special cases.

\subsubsection*{Lower Spins}

The constraints~(\ref{2.14}) on the gravitational background are necessary when the bosonic field has
spin $s\geq3$. Because $d^\m d^\n d^\r\F=0$ for a spin-2 field, the quantity $X_{\m\n\r}{}^{\a\b}$ does
not need to vanish in order for the commutator~(\ref{master}) to close. The constraints on the gravitational
background therefore boil down to:
\beq \text{For}~s=2:\qquad Y_{\m\n\r}=0,\qquad Z_{\m\n\r}=0.\eeq{2.30}
Among others, these conditions admit manifolds with a covariantly constant Ricci tensor (Ricci symmetric
spaces) reported in~\cite{Cortese:2013lda}, and in particular Einstein manifolds~\cite{Buch1,Buch2}. No
restriction on the gravitational background is imposed for $s=1$ and $s=0$.

\subsubsection*{3D Manifolds}

The Weyl tensor vanishes identically in $D=3$, and so does the tensor $X_{\m\n\r}{}^{\a\b}$. Therefore, the
necessary constraints on the gravitational background again take the form:
\beq \text{For}~s\geq2:\qquad Y_{\m\n\r}=0,\qquad Z_{\m\n\r}=0.\eeq{2.31}
The constraint on $Z_{\m\n\r}$ is tantamount to the vanishing of the Cotton tensor. In other words, it is
necessary that the 3D manifold be conformally flat.

\subsection{Electromagnetic Background}\label{sec:mbem}

Let us assume that the massive boson possesses minimal coupling to the electromagnetic (EM) background
field with an electric charge $q$. The zeroth-order deformations are obtained by the substitution:
$\de_\m\rightarrow\mD_\m$, where the covariant derivatives have commutators:
$[\mD_\m,\mD_\n]=iqF_{\m\n}$, with $F_{\m\n}$ being the background field strength. In this case,
the most generic parity-preserving deformations of the operators~(\ref{2.1}) can be written as:
\begin{equation}\label{2.32} \begin{split}
\text{Klein-Gordon}:~\bar{g}_0&=\mD^2-\bar{M}^2+iq\a F_{\m\n}u^\m d^\n+\otwob,\\
\text{Divergence}:~\bar{g}_1&=d\!\cdot\!\mD+\otwob,\\
\text{Trace}:~\bar{g}_2&=d^2+\otwob,\end{split}\end{equation}
where the deformed mass $\bar{M}^2$ and the dimensionless operator $\a$ have weight zero, and
the scale $\bar{\L}$ is larger than other mass scales in the theory. Here, the charge $q$ plays the r\^ole
of the parameter that keeps track of the deformation order (see Appendix~\ref{sec:invdeform}). The
deformations~(\ref{2.1}), of course, preserve the respective weights $w$ of the operators. Because the
deformations are smooth, $\bar{M}^2\rightarrow m^2$ in the limit of vanishing field strength.

Let us calculate the commutators between two different operators in~(\ref{2.32}). They involve the gradient
of the EM field strength, which can be decomposed as:
\beq \underbrace{\yng(1)~\otimes~\yng(1,1)}_{\text{Gradient of $F$}}~=\quad\underbrace{\yng(2,1)}_{A}
~~\oplus~~\underbrace{\yng(1)}_{V}~,\eeq{tableaux2}
where the Young diagram $\tiny{\yng(1,1,1)}$ does not contribute because of the Bianchi identity, and the other
irreducible Lorentz tensors are defined as:
\beq A_{\m\n}{}^\r~\equiv~\de_{(\m}F_{\n)}{}^\r-\left(\tfrac{1}{D-1}\right)\left[\h_{\m\n}V^\r-\d^\r_{(\m}V_{\n)}\right],
\qquad V_\n~\equiv~\de^\m F_{\m\n}.\eeq{AVdefined}
The commutators we are interested in ought to close within the given set of operators~(\ref{2.32}), so that
some deformed gauge identities to exist. We obtain (see Appendix~\ref{sec:tech2}):
\begin{equation}\label{2.36} \begin{split}
[\bar{g}_1,\bar{g}_0]&=iq\left(\a-2\right)F_{\m\n}\mD^\m d^\n-iq\a A_{\m\n\r}u^\r d^\m d^\n
+iq\left(\tfrac{\a N+(\a-1)(D-1)}{D-1}\right)d\!\cdot\!V\\
&-iq\a\left(\tfrac{1}{D-1}\right)u\!\cdot\!Vd^2-iq[\a,d\!\cdot\!\mD]F_{\m\n}u^\m d^\n+[\bar{M}^2,d\!\cdot\!\mD]
+\otwob.\end{split}\end{equation}
On the right hand side of Eq.~(\ref{2.36}), the first term  must vanish, which sets:
\beq \a=2,\eeq{setalpha}
for a $F_{\m\n}\neq0$. On the other hand, the second and third terms require that for any bosonic field of spin
$s>1$, the irreducible Lorentz tensors $A$ and $V$ vanish:
\beq A_{\m\n\r}=0,\qquad V_\m=0,\eeq{2.38}
which is tantamount to the requirement of a constant EM background: $F_{\m\n}=\text{constant}$. The remaining
problematic term is the commutator $[\bar{M}^2,d\!\cdot\!\mD]$, which can be managed by assuming again that
$\bar{M}^2$ is a polynomial function of the index operator $N$. This gives:
\beq [\bar{M}^2,d\!\cdot\!\mD]=\bar{P}(N)d\!\cdot\!\mD,\qquad [\bar{M}^2,d^2]=\bar{Q}(N)d^2,\eeq{2.39}
where $\bar{P}(N)$ and $\bar{Q}(N)$ are polynomials in $N$ of the same order. With the choices and
constraints~(\ref{setalpha})--(\ref{2.39}), the commutator~(\ref{2.36}) and the other two can be written as:
\beq [\bar{g}_1,\bar{g}_0]=\bar{P}(N)\bar{g}_1+\otwob,\qquad [\bar{g}_2,\bar{g}_0]=\bar{Q}(N)\bar{g}_2+\otwob,
\qquad [\bar{g}_1,\bar{g}_2]=\otwob.\eeq{2.40}

In view of Eqs.~(\ref{2.40}), we can identify the deformed counterparts of the commutators appearing in Eq.~(2.2)
as the following:
\beq \bar{c}_1\equiv[\bar{g}_0,\bar{g}_1]+\bar{P}(N)\bar{g}_1,\qquad \bar{c}_2\equiv[\bar{g}_2,\bar{g}_0]-\bar{Q}(N)
\bar{g}_2,\qquad \bar{c}_3\equiv[\bar{g}_1,\bar{g}_2].\eeq{2.41}
Next, we identify the deformed counterpart of the operator $j_3$ of Eq.~(\ref{2.65}); it is:
\beq \bar{j}_3\equiv\bar{g}_2\bar{c}_1+\bar{g}_1\bar{c}_2+\bar{g}_0\bar{c}_3.\eeq{2.42}
Thanks to the Jacobi identity among the deformed operators $\left\{\bar{g}_0,\bar{g}_1,\bar{g}_2\right\}$,
we can use the definitions~(\ref{2.41}) to express $\bar{j}_3$ in the following form:
\beq \bar{j}_3=\bar{c}_3\,\bar{g}_0+\left[\bar{c}_2+\bar{g}_2\bar{P}(N)+\bar{Q}(N)\bar{g}_2\right]
\bar{g}_1+\left[\bar{c}_1-\bar{P}(N)\bar{g}_1-\bar{g}_1\bar{Q}(N)\right]\bar{g}_2.\eeq{2.43}

Let us now present the deformed involutive system of equations; it is:
\beq \bar{g}_0\F=0,\qquad \bar{g}_1\F=0,\qquad \bar{g}_2\F=0.\eeq{2.44}
Assuming that the $\bar{c}_i$'s defined in Eqs.~(\ref{2.41}) vanish identically, we also have the following
gauge identities at reducibility order zero:
\beq \bar{c}_1\F=0,\qquad \bar{c}_2\F=0,\qquad \bar{c}_3\F=0,\eeq{2.45}
which holds if the $\otwob$ terms in Eqs.~(\ref{2.40}) vanish. Lacking the explicit
knowledge of similar terms in the deformed operators $\left\{\bar{g}_0,\bar{g}_1,\bar{g}_2\right\}$, the latter
condition is guaranteed if
\beq \bar{\L}\rightarrow\infty.\eeq{2.46}
We also have a desired gauge identity at reducibility order 1; it reads:
\beq \bar{j}_3\F=0,\eeq{2.47}
and holds as an on-shell identity given the relation~(\ref{2.43}). This completes our analysis of the involutive
deformation of a free massive boson in an EM background.

The consistent of dynamical equations and constraints for a free massive boson read:
\beq \left(\mD^2-\bar{M}^2+2iqF_{\m\n}u^\m d^\n\right)\F=0,\qquad d\!\cdot\!\mD\,\F=0,\qquad d^2\F=0,\eeq{2.48}
where the deformed mass $\bar{M}^2$ is assumed to be a polynomial in the index operator $N$, such that in the
limit of vanishing field strength: $\bar{M}^2\rightarrow m^2$. The consistency of this system relies on the
constraints~(\ref{2.38}) on background field strength, which mean: $F_{\m\n}=\text{constant}$. Already
found in~\cite{Cortese:2013lda}, this result holds for an arbitrary-spin\footnote{For $s=1$, because $d^\m d^\n\F=0$,
the constraint that necessarily follows from Eq.~(\ref{2.36}) is: $V_\m=0$, i.e., the EM background satisfies the
source-free Maxwell equations. For $s=0$, on the other hand, there is no constraint on the background field strength.}
boson.

\section{Massive Fermionic Fields}\label{sec:mf}

This section analyzes the Fierz-Paui system for totally-symmetric massive fermions in the operator formalism.
The starting point is the free propagation in Minkowski background, where we identify all the gauge identities
of the involutive system. Then we derive the involutive deformations in gravitational and EM backgrounds.

We use the metric convention $(-,+,\cdots,+)$. The $\g$-matrices satisfy: $\{\g^a,\g^b\}=+2\h^{ab}$, and
$\g^{a\,\dagger}=\h^{aa}\g^a$. Totally antisymmetric products of $\g$-matrices,
$\g^{a_1\cdots a_p}\equiv\g^{[a_1}\g^{a_2}\cdots\g^{a_p]}$, have unit weight.
A ``slash'' denotes a contraction with a $\gamma$-matrix, e.g., $\des=\g^a\de_a$.

\subsection{Minkowski Background}\label{sec:mfflat}

The Fierz-Pauli conditions describing a symmetric fermionic field of mass $m$ involve the Dirac, divergence
and gamma-trace operators~\cite{Rahman:2015pzl}. These operators form the set:
\beq \mathcal{F}=\left\{f_0,g_1,f_1\right\},\eeq{3.1}
where again a subscript gives the negative weight ($-w$) corresponding to an operator. Table~2 summarizes
the various properties of these operators.
\begin{table}[ht]
\caption{Operators in Fermionic Fierz-Pauli System}
\vspace{6pt}
\centering
\begin{tabular}{c c c c c}
\hline\hline
Operator~~~&~~~Symbol~~~&~~~Definition~~&~~Weight $(w)$~~&~~Derivative Order $(k)$\\ [0.5ex]
\hline
Dirac & $f_0$ & $\des-m$ & $~~0$ & 1\\
Divergence & $g_1$ & $\div$ & $-1$ & 1\\
Gamma-Trace & $f_1$ & $\ds$ & $-1$ & 0\\
\hline\hline
\end{tabular}
\end{table}
\vspace{6pt}

\noindent
We will be interested in the graded commutators between two \emph{different} operators: $[f_0,g_1]$,
$[g_1,f_1]$ and $\{f_1,f_0\}$. The first two commutators vanish, while the last one is given by:
\beq \{f_1,f_0\}=2g_1-2mf_1,\eeq{3.3}
which closes within the given set $\mathcal{F}$. Let us now define the following operators:
\bea &h_1\equiv[f_0,g_1],\qquad h_2\equiv[g_1,f_1],\qquad h_1'\equiv\{f_1,f_0\}-2g_1+2mf_1,&\label{3.3.1}\\
&j_2\equiv f_1h_1-\left(f_0+2m\right)h_2+g_1h_1'.&\eea{3.3.2}
Because the operators $\left\{f_0,g_1,f_1\right\}$ are linear, we have the graded Jacobi identity:
\beq \{f_1,[f_0,g_1]\}-\{f_0,[g_1,f_1]\}+[g_1,\{f_1,f_0\}]=0,\eeq{3.4}
which enables us to rewrite the operator $j_2$, defined in Eq.~(\ref{3.3.2}), as:
\beq j_2=h_2f_0+h_1'g_1-h_1f_1.\eeq{3.4.1}

The Fierz-Pauli equations comprise an involutive system of differential equations~\cite{Inv-book}:
\beq f_0\Ps=0,\qquad g_1\Ps=0,\qquad f_1\Ps=0.\eeq{3.5}
The mutual compatibility of the equations~(\ref{3.5}) is encoded in the gauge identities:
\beq h_1\Ps=0,\qquad h_2\Ps=0,\qquad h_1'\Ps=0,\eeq{3.6}
which follow directly from the graded commutators of the operators in $\mathcal{F}$. Moreover, because
of the relation~(\ref{3.4.1}), we have the following \emph{on-shell} identity:
\beq j_2\Ps=0,\eeq{3.7}
which is a gauge identity at reducibility order 1, implying that the gauge identities~(\ref{3.6}) are not
irreducible. This completes the list of all possible gauge identities of our system.

Note that the system~(\ref{3.5})--(\ref{3.7}) of involutive equations and gauge identities is of the type
considered in Appendix~\ref{sec:nogauge}. In order to check its absolute compatibility and count the DoF's,
we first give the number of equations at order $k$ in \emph{space-time derivatives}:
\beq t_k=\left[\d_k^1{D+n-1\choose n}+\d_k^1{D+n-2\choose n-1}+\d_k^0{D+n-2\choose n-1}\right]2^{[D]/2},\eeq{3.8}
where $n$ is the rank of the symmetric fermion, and $D$ is the space-time dimensionality.
We also have the count of $\mathcal{O}(k)$ gauge identities at reducibility order $j$, given by:
\beq l_{k,\,j}=\left[\d_k^2\d_j^0{D+n-2\choose n-1}+\d_k^1\d_j^0{D+n-2\choose n-1}
+\left(\d_k^1\d_j^0+\d_k^2\d_j^1\right){D+n-3\choose n-2}\right]2^{[D]/2}.\eeq{3.9}
Given the total number of field variables $f={D+n-1\choose n}2^{[D]/2}$, we find from Eq.~(\ref{A17})
that $\mc=0$, i.e., the fermionic Fierz-Pauli system is absolutely compatible. The count of physical DoF's
per space-time point is given by Eq.~(\ref{A18}):
\beq \mathfrak{D}_f(n)={D+n-3\choose n}\,2^{[D-2]/2},\eeq{3.12}
which is the number of propagating DoF's of a massive spin-$\left(n+\tfrac{1}{2}\right)$ fermion~\cite{Rahman:2015pzl}.

\subsection{Gravitational Background}\label{sec:mfgrav}

The free propagation of a massive fermion in a gravitational background can be analyzed by applying the involutive
deformation method to the flat-space system we just described. In accordance with Appendix~\ref{sec:invdeform},
the substitution of ordinary derivatives by covariant ones, $\de_\m\rightarrow\nb_\m$, gives the zeroth-order
deformations, while linear terms in the background curvature comprise the first-order ones, etc. The deformations
of the operators~(\ref{3.1}) ought to preserve the respective weights $w$; they can be written as:
\begin{equation}\label{3.13}\begin{split}
\text{Dirac}:~\hat{f}_0&=\nbs-M+\oone,\\
\text{Divergence}:~\hat{g}_1&=\ddiv+\oone,\\
\text{Gamma-Trace}:~\hat{f}_1&=\ds+\otwo,\end{split}\end{equation}
where the deformed mass $M$ has weight $w=0$, and $\L$ is some mass scale larger than other scales in the theory.
The deformations~(\ref{3.13}) are assumed to be smooth, so that in the limit of zero curvature: $M\rightarrow m$.
Here, the book-keeping parameter indicating the deformation order (see Appendix~\ref{sec:invdeform}) is implicit.

We will now compute the graded commutators between two different operators in~(\ref{3.13}). The details of the
computations are given in Appendix~\ref{sec:tech1}. We must ensure that these commutators close within the given
set of operators, so that some deformed versions of the gauge identities exist at all. An explicit computation
leads us to the following result:
\begin{equation}\label{3.14}\begin{split}
[\hat{f}_0,\hat{g}_1]&=W_{\m\n\r\s}\g^\m u^\r d^\n d^\s+\left(\tfrac{1}{D-2}\right)\left[\slashed{u}\,
S_{\m\n}d^\m d^\n-\left(\tfrac{2N+D-2}{2}\right)S_{\m\n}\g^\m d^\n\right]-[M,\ddiv]\\
&+\left(\tfrac{1}{D-2}\right)\left[S_{\m\n}\g^\m u^\n\ds-S_{\m\n}u^\m d^\n\right]\ds
+\tfrac{1}{D(D-1)}\,R\left[\slashed{u}\,\ds-\left(\tfrac{2N+D-1}{2}\right)\right]\ds+\oone.
\end{split}\end{equation}
From the first
line of Eq.~(\ref{3.14}) it is clear that, for an arbitrary-spin field, the gravitational background
is required to fulfill the following conditions:
\beq W_{\m\n\r\s}=0,\qquad S_{\m\n}=0.\eeq{3.15}
In other words, the background manifold must be a conformally flat as well as an Einstein one. This is tantamount
to the requirement of a maximally symmetric space, for which Eqs.~(\ref{C2}) apply.
We also need to deal with the commutator $[M,\ddiv]$. In order to do so, let us assume that the deformed mass $M$
is a linear function of the index operator $N$:
\beq M=m+\m\left(N+\d\right),\eeq{3.16}
where $\d$ is a numerical constant, and $\m$ a constant mass parameter that vanishes in the zero curvature limit.
Again, the justification of such an assumption is postponed until we consider the massless case. The
constraints~(\ref{3.15}) and the choice~(\ref{3.16}) reduce the commutator~(\ref{3.14}) to a desired form.
In an AdS space of radius $L$, one obtains:
\beq [\hat{f}_0,\hat{g}_1]=\m\hat{g}_1-\rad\left[\slashed{u}\,\ds-\left(\tfrac{2N+D-1}{2}\right)\right]
\hat{f}_1+\oone.\eeq{3.17}
The other graded commutators, on the other hand, are simpler to compute. They read:
\beq [\hat{g}_1,\hat{f}_1]=\oone,\qquad \{\hat{f}_1,\hat{f}_0\}=2\hat{g}_1-\left(2M+\m\right)\hat{f}_1+\oone.
\eeq{3.18}

With the graded commutation relations~(\ref{3.17})--(\ref{3.18}), we can now identify the deformed counterparts
of the operators~(\ref{3.3.1}); they are given by:
\bea &\hat{h}_1\equiv[\hat{f}_0,\hat{g}_1]-\m\hat{g}_1+\rad\left[\slashed{u}\,\ds
-\left(\tfrac{2N+D-1}{2}\right)\right]\hat{f}_1,&\nonumber\\
&\hat{h}_1'\equiv\{\hat{f}_1,\hat{f}_0\}-2\hat{g}_1+\left(2M+\m\right)\hat{f}_1,&\label{3.19}\\
&\hat{h}_2\equiv[\hat{g}_1,\hat{f}_1].&\nonumber\eea{xxyyzzz}
We also identify the deformed counterpart of the operator $j_2$ in Eq.~(\ref{3.3.2}) with:
\beq \hat{j}_2\equiv\hat{f}_1\hat{h}_1-\left(\hat{f}_0+2(M+\m)\right)\hat{h}_2+\hat{g}_1\hat{h}_1'.\eeq{3.20}
The graded Jacobi identity involving the operators $\left\{\hat{f}_0,\hat{g}_1,\hat{f}_1\right\}$, however, gives:
\beq \hat{j}_2=\hat{h}_2\hat{f}_0+\hat{h}_1'\hat{g}_1
-\left[\hat{h}_1-\rad\left(\{\hat{f}_1,\slashed{u}\,\ds-N\}-(D-1)\hat{f}_1\right)\right]\hat{f}_1.\eeq{3.21}

At this stage, we are ready to present the deformed involutive system along with all the gauge identities.
The dynamical equations and constraints read:
\beq \hat{f_0}\Ps=0,\qquad \hat{g}_1\Ps=0,\qquad \hat{f}_1\Ps=0,\eeq{3.22}
while the gauge identities at reducibility order 0 are:
\beq \hat{h}_1\Ps=0,\qquad \hat{h}_2\Ps=0,\qquad \hat{h}_1'\Ps=0,\eeq{3.23}
which follow from the graded commutators~(\ref{3.17})--(\ref{3.18}) provided that the $\oone$-terms appearing
therein vanish. The latter conditions can be ensured by taking:
\beq \L\rightarrow\infty.\eeq{3.24}
Furthermore, the relation~(\ref{3.21}) gives rise to the following \emph{on-shell} identity:
\beq \hat{j}_2\Ps=0,\eeq{3.25}
which is the desired gauge identity at reducibility order 1. This completes our analysis.

To summarize, the involutive system of equations for a massive fermion reads:
\beq \left(\nbs-M\right)\Ps=0,\qquad \ddiv\Ps=0,\qquad\ds\,\Ps=0,\eeq{3.26}
where the deformed mass $M$ is assumed to be of the form~(\ref{3.16}). For a fermion of arbitrary
spin, this system is consistent in $D\geq3$ when the gravitational background is a maximally symmetric
space. The constraints are weaker for lower-spin fields. In particular, as already noted
in~\cite{Buchbinder:2010gp,Rahman:2011ik}, a spin-$\tfrac{3}{2}$ massive fermion can be consistently
described in an Einstein space ($S_{\m\n}=0$). This can easily be seen from Eq.~(\ref{3.14}) given that
in this case $d^\m d^\n\Ps=0$. No such constraints on the gravitational background appear for
$s=\tfrac{1}{2}$.

\subsection{Electromagnetic Background}\label{sec:mfem}

We assume that the massive fermion has a nonzero electric charge $q$, which defines its minimal coupling to the
EM background. As usual, the zeroth-order deformations are obtained by the substitution: $\de_\m\rightarrow\mD_\m$.
So, the deformations of the operators~(\ref{3.1}) are:
\begin{equation}\label{3.27}\begin{split}
\text{Dirac}:~\bar{f}_0&=\slashed{\mD}-m+\mathcal{A},\\
\text{Divergence}:~\bar{g}_1&=d\!\cdot\!\mD+\mathcal{B},\\
\text{Gamma-Trace}:~\bar{f}_1&=\ds+\mathcal{C},\end{split}\end{equation}
where $\mathcal{A}$, $\mathcal{B}$ and $\mathcal{C}$ contain all the higher-order deformations that are assumed to
be smooth and parity preserving. Note that the deformation order is controlled by the charge $q$.

Given the formal expressions~(\ref{3.27}), one can write down the graded commutators between two different operators.
They read:
\bea &[\bar{f}_0,\bar{g}_1]=iqF_{\m\n}\g^\m d^\n+\big(\des\mathcal{B}-\div\mathcal{A}\big)
+\big([\g^\m,\mathcal{B}]-[d^\m,\mathcal{A}]\big)\mD_\m+[\mathcal{A},\mathcal{B}],&\label{3.28}\\
&[\bar{g}_1,\bar{f}_1]=[\mathcal{B},\ds]-\div\,\mathcal{C}-[d^\m,\mathcal{C}]\mD_\m+[\mathcal{B},\mathcal{C}],
\label{3.29}\\&\{\bar{f}_1,\bar{f}_0\}=2\bar{g}_1-2m\bar{f}_1+\{\ds,\mathcal{A}\}
-2\mathcal{B}+2m\mathcal{C}+\{\mathcal{C},\mathcal{A}\}.&\eea{3.30}
These commutators ought to close within the set of operators~(\ref{3.27}). The $F_{\m\n}\g^\m d^\n$-term
in Eq.~(\ref{3.28}) requires that the non-minimal couplings be present, i.e., the terms $\mathcal{A}$,
and $\mathcal{B}$ cannot both be zero because otherwise the commutator $[\bar{f}_0,\bar{g}_1]$
does not close.

It is difficult to find the general solution for $\mathcal{A}$, $\mathcal{B}$ and $\mathcal{C}$ for generic
spin. In order to proceed, we will therefore make some simplifying assumptions. First, let us assume that
\beq \mathcal{C}=0.\eeq{3.31}
In other words, the $\g$-trace operator does not undergo any deformation at order one or higher. This can be
justified by noting that all the known consistent models of charged massive higher-spin fields enjoy this
property~\cite{AN,Klishevich,PRS,PR3half}. Moreover, such deformations may not show up even in a gravitational
background, as we just saw. Next, we spell out the non-minimal deformation of the
Dirac operator (see Appendix~\ref{sec:tech2}):
\beq \mathcal{A}=iq\left(a_{+}F^{+}_{\m\n}+a_{-}F^{-}_{\m\n}\right)u^\m d^\n
+iq\left(a_0 F_{\r\s}\g^{\r\s}+\cdots\right)+\otwoq,\eeq{3.32}
where $F^{\pm}_{\m\n}\equiv F^{\m\n}\pm\tfrac{1}{2}\g^{\m\n\r\s}F_{\r\s}$, the $a_{\pm}$ and $a_0$ are
operators of weight $w=0$ and mass dimension $-1$, and the ellipses stand for terms containing derivatives
of the field strength. Similarly, we can write down the non-minimal deformation of the divergence:
\beq \mathcal{B}=iq\left(b_0 F_{\m\n}\g^\m d^\n+\cdots\right)+\otwoq,\eeq{3.33}
with $b_0$ being a weight-$0$ operator of dimension $-1$, and the ellipses containing derivatives of the
field strength. Given Eqs.~(\ref{3.31})--(\ref{3.33}), one can compute the graded commutators up to $\ooneq$,
as in Appendix~\ref{sec:tech2}. For spin $s\geq\tfrac{3}{2}$, the cancellation of the offending $\ooneq$ terms
obstructing the closure of the commutators~(\ref{com1}) and~(\ref{ancom}) requires that:
\bea &1-m\left(a_{+}-a_{-}+2b_0\right)=0,&\nonumber\\
    &a_{-}-b_0=0,&\label{lin}\\
    &(D-4)a_{+}-(D-2)a_{-}+4a_0+2b_0=0,\nonumber\eea{naidgdjsk}
which can be solved, with the introduction of a single free parameter $\e$, as:
\beq a_{\pm}=\tfrac{1}{2}\left(1\pm\e\right)m^{-1},\qquad a_0=-\left(\tfrac{D-4}{4}\right)\e m^{-1},\qquad
b_0=\tfrac{1}{2}\left(1-\e\right)m^{-1}.\eeq{3.34}
Moreover, the irreducible Lorentz tensors $A_{\m\n}$ and $V_\m$ (see Eq.~(\ref{AVdefined})) must vanish, i.e.,
\beq F_{\m\n}=\text{constant}.\eeq{3.35}
With these choices and constraints, the graded commutators~(\ref{3.28})--(\ref{3.30}) reduce to:
\begin{equation}\label{3.36}\begin{split}
[\bar{f}_0,\bar{g}_1]&=(iq/m)F_{\m\n}\left[-\g^\m d^\n\bar{f}_0+\tfrac{1}{2}\e\left(\g^{\m\n}\bar{g}_1
+2\g^\m\mD^\n\bar{f}_1-\g^{\m\n}\slashed{\mD}\bar{f}_1\right)\right]+\otwoq,\\
[\bar{g}_1,\bar{f}_1]&=(iq/m)\left(1-\e\right)F_{\m\n}\g^\m d^\n\bar{f}_1+\otwoq,\\
\{\bar{f}_1,\bar{f}_0\}&=2\bar{g}_1-2m\bar{f}_1+(iq/m)F_{\m\n}\left[2u^\m d^\n
+\tfrac{1}{2}\e\g^{\m\n}\right]\bar{f}_1+\otwoq.\end{split}\end{equation}

We are now ready to identify the deformations of the operators $\{h_1,h_2,h_1'\}$ given in Eq.~(\ref{3.3.1}).
Up to $\otwoq$ correction terms, they are:
\bea
\bar{h}_1&\equiv&[\bar{f}_0,\bar{g}_1]+(iq/m)F_{\m\n}\left[\g^\m d^\n\bar{f}_0-\tfrac{1}{2}\e\left(\g^{\m\n}\bar{g}_1
+2\g^\m\mD^\n\bar{f}_1-\g^{\m\n}\slashed{\mD}\bar{f}_1\right)\right],\nonumber\\
\bar{h}_2&\equiv&[\bar{g}_1,\bar{f}_1]-(iq/m)\left(1-\e\right)F_{\m\n}\g^\m d^\n\bar{f}_1,\label{3.37}\\
\bar{h}_1'&\equiv&\{\bar{f}_1,\bar{f}_0\}-2\bar{g}_1+2m\bar{f}_1-(iq/m)F_{\m\n}\left[2u^\m d^\n
+\tfrac{1}{2}\e\g^{\m\n}\right]\bar{f}_1.\nonumber\eea{amnmsjk}
We also identify the deformed counterpart of the operator $j_2$ in Eq.~(\ref{3.3.2}); it is:
\beq \bar{j}_2\equiv\bar{f}_1\bar{h}_1-\left(\bar{f}_0+2m\right)\bar{h}_2+\bar{g}_1\bar{h}_1'.\eeq{3.38}
Thanks to the graded Jacobi identity involving the operators $\left\{\bar{f}_0,\bar{g}_1,\bar{f}_1\right\}$,
one can use the definitions~(\ref{3.37}) to rewrite $\bar{j}_2$ in the following form:
\beq \bar{j}_2=\bar{O}_2\,\bar{f}_0+\bar{O}_1'\,\bar{g}_1-\bar{O}_1\,\bar{f}_1+\otwoq,\eeq{3.39}
where the explicit expressions of the operators $\bar{O}_2$, $\bar{O}_1'$ and $\bar{O}_1$ are given in
Eqs.~(\ref{C40}).

Our deformed involutive system consists of the dynamical equations and constraints:
\beq \bar{f_0}\Ps=0,\qquad \bar{g}_1\Ps=0,\qquad \bar{f}_1\Ps=0.\eeq{3.40}
The required gauge identities are valid up to $\ooneq$. At reducibility order 0, they read:
\beq \bar{h}_1\Ps=\otwoq,\qquad \bar{h}_2\Ps=\otwoq,\qquad \bar{h}_1'\Ps=\otwoq,\eeq{3.41}
thanks to the graded commutators~(\ref{3.36}). At reducibility order 1, the gauge identity is:
\beq \bar{j}_2\Ps=\otwoq,\eeq{3.42}
which is an \emph{on-shell} identity that follows from the relation~(\ref{3.39}).

Therefore, a free massive fermion of spin $s\geq\tfrac{3}{2}$ in an EM background is described, up to $\ooneq$,
by the following one-parameter family of an involutive system of equations:
\bea &\left\{\slashed{\mD}\!-\!m\!+\!(iq/m)\!\left[\left(F_{\m\n}\!+\!\tfrac{1}{2}\e\g_{\m\n\r\s}F^{\r\s}\right)\!
u^\m d^\n\!-\!\left(\tfrac{D-4}{4}\right)\!\e F_{\m\n}\g^{\m\n}\right]\!+\!\otwoq\right\}\Ps=0,&\label{3.44}\\
&\left\{d\!\cdot\!\mD+\tfrac{1}{2}(iq/m)(1-\e)F_{\m\n}\g^\m d^\n+\otwoq\right\}\Ps=0,\qquad\ds\,\Ps=0,\eea{3.45}
given that the background is a constant one: $F_{\m\n}=\text{constant}$. In principle, one can proceed order by order
in the parameter $q$ to find the higher-order deformations. However, it is  not clear at all whether a consistent
deformation up to all order exists for arbitrary spin. The only known example of an all-order
solution is for $s=\tfrac{3}{2}$ in $D=4$~\cite{PR3half}, to which\footnote{It has the Dirac equation: $\left[\slashed{\mD}-m+m\left(B^+_{\m\n}-B_\m{}^\r B_{\r\n}+\tfrac{1}{4}\h_{\m\n}\text{Tr}B^2\right)u^\m d^\n\right]\Ps=0$,
plus constraints: $\left(d\!\cdot\!\mD+\tfrac{1}{2}mB_\m{}^\r B_{\r\n}\g^\m d^\n\right)\Ps=0$,~and~$\ds\,\Ps=0$, where
$B_{\m\n}=(iq/m^2)F_{\m\n}+\tfrac{1}{4}\text{Tr}B^2B_{\m\n}-\tfrac{1}{4}\text{Tr}(B\tilde{B})\tilde{B}_{\m\n}$.} our
$\ooneq$-results~(\ref{3.44})--(\ref{3.45}) agree, with the parameter choice of $\e=1$.

\section{Massless Bosonic Fields}\label{sec:gb}

In this section, we consider the zero mass limit of the involutive system of a higher-spin massive boson. As we will
see, in the massless limit the flat-space involutive system~(\ref{2.5}) acquires a gauge symmetry, whose gauge parameter
itself is governed by the same kind of involutive system. In other words, we obtain the description of a higher-spin
gauge boson in the transverse-traceless gauge. Given the discussion of Appendix~\ref{sec:constgauge}, we then confirm
that the involutive system of a gauge boson describes the correct number of physical DoF's. Armed with this formulation,
we then study the consistent free propagation of massless bosons in nontrivial backgrounds.

\subsection{Minkowski Background}\label{sec:gbflat}

For the massive spin-$s$ boson $\F$ of Eqs.~(\ref{2.5}), let us consider the following transformation:
\beq \d\F=g_{-1}\l,\qquad \l=\tfrac{1}{(s-1)!}\,\l_{\m_1\cdots\m_{s-1}}(x)u^{\m_1}\cdots u^{\m_{s-1}},\eeq{4.1}
where we have introduced the symmetrized gradient operator $g_{-1}$, defined as:
\beq\text{Symmetrized Gradient:}\quad g_{-1}~\equiv~\grad,\qquad \text{with}\quad [N,g_{-1}]=g_{-1}.\eeq{4.2}
We take note of the following commutation relations for the symmetrized gradient:
\beq [g_{0},g_{-1}]=0,\qquad[g_{1},g_{-1}]=g_{0}+m^2,\qquad[g_{2},g_{-1}]=2g_{1},\eeq{4.3}
to find that the left-hand sides of the involutive equations~(\ref{2.5}) transform as:
\bea \d(g_0\F)&=&g_{-1}(g_0\l),\nonumber\\
\d(g_1\F)&=&g_{-1}(g_1\l)+(g_0+m^2)\l,\label{4.4}\\
\d(g_2\F)&=&g_{-1}(g_2\l)+2g_1\l.\nonumber\eea{ekxsnsl}

We would like to see when, if at all, transformations of the type~(\ref{4.1}) may become a symmetry
of the Fierz-Pauli involutive system~(\ref{2.5}). With this end in view, let us first impose that
$\l$ itself be governed by the following involutive set of equations:
\beq g_0\l=0,\qquad g_1\l=0,\qquad g_2\l=0.\eeq{4.5}
Then, the right-hand sides of Eqs.~(\ref{4.4}) vanish if:
\beq m^2\l=0.\eeq{4.500}
Therefore, a nontrivial gauge symmetry emerges in the massless limit: $m^2\rightarrow0$.

In other words, the involutive system of a massless boson enjoys a gauge symmetry~(\ref{4.1}), where the gauge
parameter satisfies Eqs.~(\ref{4.5}) with zero mass. The Klein-Gordon operator reduces in this case to the
d'Alembertian operator, denoted as:
\beq \text{d'Alembertian:}\quad \dal~\equiv~\bbox=\lim_{m^2\rightarrow0}g_0,\qquad \text{with}\quad [N,\dal]=0.
\eeq{4.6}
The operators relevant for the massless case are the massless cousins of~(\ref{2.1}) and the symmetrized gradient,
which we collect in the following set:
\beq \underline{\mathcal{G}}=\{\dal,g_1,g_2,g_{-1}\}.\eeq{4.8}
Notice that the massless counterparts of the commutators~(\ref{4.3}) are:
\beq [\dal,g_{-1}]=0,\qquad[g_{1},g_{-1}]=\dal,\qquad[g_{2},g_{-1}]=2g_{1},\eeq{4.7}
which close completely within the set $\underline{\mathcal{G}}\backslash\left\{g_{-1}\right\}$. This fact plays
a crucial r\^ole in the existence of transverse-traceless gauge symmetry. It is the closure of the
commutators~(\ref{4.7}) that ensures gauge invariance, which in turn controls the DoF count, as we will now see.

In order to make the DoF count, let us note that the gauge field $\F$ and the gauge parameter $\l$ are
both governed by the same set of involutive equations, which is:
\beq \dal\doub=0,\qquad g_1\doub=0,\qquad g_2\doub=0.\eeq{4.9}
It is easy to see from Section~\ref{sec:mbflat} that the zero mass limit does not hurt the involutive
structure of the Fierz-Pauli system~(\ref{2.5}). Neither does it alter the DoF count~(\ref{2.12}). In
this case, however, the aforementioned count is a naive one because of  the emergence of gauge symmetry.
This is precisely the circumstances under which the analysis of Appendix~\ref{sec:constgauge} may apply.
From formula~(\ref{A24}), it is easy to write down the number of physical DoF for a spin-$s$ gauge field;
it is simply the difference between the DoF count of a massive spin-$s$ boson and that of a massive
spin-$(s-1)$ boson:
\beq \mathfrak{D}_b^{(0)}(s)=\mathfrak{D}_b(s)-\mathfrak{D}_b(s-1).\eeq{4.11}
Then, it follows directly from the DoF count formula~(\ref{2.12}) that
\beq \mathfrak{D}_b^{(0)}(s)=2\,{D-5+s\choose s-1}+{D-5+s\choose s},\eeq{4.12}
which is the correct number of propagating DoF's for a massless spin-$s$ boson~\cite{Rahman:2015pzl}.

\subsection{Gravitational Background}\label{sec:gbgrav}

In a gravitational background, we would like to find the deformed counterparts of the operators~(\ref{4.8}).
The massless limits of the deformed operators in Eqs.~(\ref{2.28})--(\ref{2.29}), augmented by the
deformed symmetrized gradient $\hat{g}_{-1}$ give following set:
\beq \underline{\hat{\mathcal{G}}}=\{\dalh,\hat{g}_1,\hat{g}_2,\hat{g}_{-1}\}.\eeq{4.13}
This includes the deformed d'Alembertian operator:
\beq \dalh=\dbbox-M_0^2+R_{\m\n\r\s}u^\m u^\r d^\n d^\s-R_{\m\n}u^\m d^\n+\tfrac{2(N-1)(N+D-2)}{(D-1)(D+2)}\,R,
\eeq{4.14}
where, we recall from the mass ansatz~(\ref{2.16}) that,
\beq M_0^2=\m^2\left(N^2+\b N+\g\right),\eeq{4.15}
with $\m$ being a constant mass parameter that vanishes in the zero curvature limit, and $\b$ and $\g$ numerical
constants. We also have the deformed divergence and trace operators:
\beq \hat{g}_1=\ddiv,\qquad \hat{g}_2=d^2.\eeq{4.16}
Last but not the least, we have the deformed symmetrized gradient. To write this,
let us recall from Eq.~(\ref{2.26}) that we choose to stay in a parametric regime where the suppression scale $\L$
of higher-dimensional operators is taken to infinity. This allows us to drop all the possible non-minimal terms to
$\hat{g}_{-1}$, and instead identify it as a zeroth order deformation:
\beq \hat{g}_{-1}=\dgrad.\eeq{4.17}

The involutive system of a spin-$s$ massless boson $\F$ is given simply by the massless limits of
Eqs.~(\ref{2.28})--(\ref{2.29}), i.e., through the deformed operators~(\ref{4.14})--(\ref{4.16}), as:
\beq \dalh\F=0,\qquad \hat{g}_1\F=0,\qquad \hat{g}_2\F=0.\eeq{4.18}
The spin-$(s-1)$ gauge parameter $\l$, on the other hand, is governed by a similar system:
\beq \dalh'\l=0,\qquad \hat{g}_1\l=0,\qquad \hat{g}_2\l=0.\eeq{4.19}
In order not to ruin the involutive structure of Eqs.~(\ref{4.19}), the deformed d'Alembertian $\dalh'$
acting on the gauge parameter may differ from $\dalh$ only in the mass-like term:
\beq \dalh'=\dalh+M_0^2-M_0^{\prime\,2},\eeq{4.20}
where $M_0^{\prime\,2}$ is some quadratic polynomial in $N$, in accordance with the ansatz~(\ref{2.16}).

We now consider gauge transformations of the form: $\d\F=\hat{g}_{-1}\l$, and find the variations of
the left-hand sides of Eqs.~(\ref{4.18}); they are given by:
\begin{equation}\label{4.21}\begin{split}
\d(\dalh\F)&~=~[\dalh,\hat{g}_{-1}]\l+\hat{g}_{-1}(\dalh\l)~=~[\dalh,\hat{g}_{-1}]\l
+\hat{g}_{-1}\left(M_0^{\prime\,2}-M_0^2\right)\l,\\
\d(\hat{g}_1\F)&~=~[\hat{g}_1,\hat{g}_{-1}]\l+\hat{g}_{-1}(\hat{g}_1\l)~=~[\hat{g}_1,\hat{g}_{-1}]\l,\\
\d(\hat{g}_2\F)&~=~[\hat{g}_2,\hat{g}_{-1}]\l+\hat{g}_{-1}(\hat{g}_2\l)~=~[\hat{g}_2,\hat{g}_{-1}]\l,
\end{split}\end{equation}
where the right-hand sides are obtained by making use of Eqs.~(\ref{4.19})--(\ref{4.20}). In order to see
how gauge invariance can be restored in a gravitational background, we therefore need the commutators of
$\hat{g}_{-1}$ with the other three operators in~(\ref{4.13}). The commutators with $\hat{g}_1$ and $\hat{g}_2$
are rather easy to compute; they can be written as:
\beq [\hat{g}_1,\hat{g}_{-1}]=\dalh'+\mathcal{X}_0,\qquad [\hat{g}_2,\hat{g}_{-1}]=2\hat{g}_1,\eeq{4.22}
where the weight-$0$ operator $\mathcal{X}_0$ is explicitly given in Eq.~(\ref{C9}). In view of
Eqs.~(\ref{4.19}), the necessary and sufficient conditions for the vanishing of $\d(\hat{g}_1\F)$ and
$\d(\hat{g}_2\F)$, i.e., for the gauge invariance of the transverse-traceless conditions amount to:
\beq \mathcal{X}_0\l=0.\eeq{4.23}
Now, using the decomposition formula~(\ref{C1}), it is possible to write:
\beq \mathcal{X}_0=-2W_{\m\n\r\s}u^\m u^\r d^\n d^\s+\left(\tfrac{2}{D-2}\right)S_{\m\n}
\left[(2N+D)u^\m d^\n-u^2d^\m d^\n-u^\m u^\n d^2\right]+\cdots,\eeq{4.24}
where the ellipses contain neither of the irreducible tensors $W_{\m\n\r\s}$ and $S_{\m\n}$. By inspection,
it is clear that in order for Eq.~(\ref{4.23}) to hold, for arbitrary spin $s>2$, the gravitational background
is required to be conformally flat as well as Einsteinian:
\beq W_{\m\n\r\s}=0,\qquad S_{\m\n}=0.\eeq{4.25}
In other words, the background must be a maximally symmetric space\footnote{Fulfilled automatically by any maximally
symmetric space, the constraints~(\ref{2.14}) are indeed weaker.}. Then, one can make use of Eq.~(\ref{C2}) to find
the following simple expression:
\beq \mathcal{X}_0L^2=-u^2\hat{g}_2+M_0^{\prime\,2}L^2-(2N+D)(N+D-2)/(1+\tfrac{1}{2}D).
\eeq{4.26}
In order for Eq.~(\ref{4.23}) to be fulfilled, the following identification must be made:
\beq M_0^{\prime\,2}L^2=(2N+D)(N+D-2)/(1+\tfrac{1}{2}D),\eeq{4.27}
which gives a justification to the mass ansatz~(\ref{2.16}). The constraints~(\ref{4.25}) and the parameter
choice~(\ref{4.27}) ensure the gauge invariance of the transverse-traceless conditions.

Next, we consider the gauge symmetry of the dynamical equation, for which we need the commutator
$[\dalh,\hat{g}_{-1}]$. This can be computed easily by taking the hermitian conjugate\footnote{In this regard,
the hermitian conjugation is implemented by: $u^\dagger_\m=d_\m$ and $d^\dagger_\m=u_\m$. Indeed one has:
$[d_\m,u_\n]=[d_\m,d^\dagger_\n]=g_{\m\n}$, which allows for interpretation in terms of creation and annihilation
operators.} of Eq.~(\ref{2.18}) in the limit $\L\rightarrow\infty$ and $m^2\rightarrow0$. Thus, we obtain:
\beq [\dalh,\hat{g}_{-1}]=
-\hat{g}_{-1}\left[\m^2\left(2N+\b+1\right)+\tfrac{1}{L^2}D(2N+D-2)/(1+\tfrac{1}{2}D)\right],\eeq{4.28}
given the constraint of maximally symmetric background. Now, let us take the first equation
of~(\ref{4.21}), and plug the expressions~(\ref{4.15}),~(\ref{4.27}) and~(\ref{4.28}) in it to write:
\beq \d(\dalh\F)=-\tfrac{1}{L^2}\,\hat{g}_{-1}\left(\d_2N^2+\d_1N+\d_0\right)\l,\eeq{4.29}
where the numerical coefficients $\d_2$, $\d_1$ and $\d_0$ are given by:
\beq \d_2=\m^2L^2-\tfrac{4}{D+2},\qquad \d_1=(\b+2)\m^2L^2-\tfrac{2(D-4)}{D-2},\qquad \d_0=(\b+\g+1)\m^2L^2.
\eeq{4.30}
Each of these coefficients must be zero since otherwise the right-hand side of Eq.~(\ref{4.29}) does not vanish.
This leads to a unique solution for the parameters $\m^2$, $\b$ and $\g$, which can be reexpressed through a
solution for the mass-like term, as:
\beq M_0^2L^2=(N-1)(2N+D-6)/(1+\tfrac{1}{2}D).\eeq{4.31}
This again justifies the mass ansatz~(\ref{2.16}). For the massive case$-$as long as the involutive structure of
the system is concerned$-$any arbitrary polynomial in the index operator $N$ would qualify as the deformed mass.
Only in the massless limit does one see why this ought to be a quadratic polynomial in $N$. Given the
constraints~(\ref{4.25}), and the expressions~(\ref{4.27}) and~(\ref{4.31}), the deformed d'Alembertians~(\ref{4.14})
and~(\ref{4.20}) reduce to:
\begin{equation}\label{4.32}\begin{split}
\dalh&=\nb^2-\mm_0^2,\qquad~\mathfrak{m}_0^2L^2\equiv(N-2)(N+D-3)-N,\\
\dalh'&=\nb^2-\mm_0^{\prime\,2},\qquad \mathfrak{m}_0^{\prime\,2}L^2\equiv N(N+D-1)-N.
\end{split}\end{equation}

Now we are ready to present our gauge invariant involutive system. The transformation of the massless spin-$s$ field
$\F$ is given in terms of a spin-$(s-1)$ gauge parameter $\l$, as $\d\F=\dgrad\l$.  They are governed by their
respective involutive systems:
\begin{equation}\label{4.33}\begin{split}
\left(\nb^2-\mathfrak{m}_0^2\right)\F&=0,\qquad \ddiv\F=0,\qquad d^2\F=0,\\
\left(\nb^2-\mathfrak{m}_0^{\prime\,2}\right)\l&=0,\qquad \ddiv\l=0,\qquad d^2\l=0,
\end{split}\end{equation}
with the mass-like terms given by Eqs.~(\ref{4.32}). This system holds good in $D\geq3$ for spin $s>2$ only in maximally
symmetric spaces. The lower-spin case is considered below.

\subsubsection*{Lower Spins}

The constraints~(\ref{4.25}) on the gravitational background are necessary only for gauge bosons with spin $s\geq3$.
The gauge parameter in the spin-$2$ case satisfies: $d^\m d^\n\l=0$, and therefore the Weyl tensor does not need to
vanish in Eq.~(\ref{4.24}) for a field with $s=2$. The necessary constraint in this case turns out to be:
\beq \text{For}~s=2:\qquad S_{\m\n}=0.\eeq{4.34}
In other words, the gravitational background must be an Einstein manifold. Note that the conditions~(\ref{2.30})
in the massive case automatically holds for such a background. The system is still described by Eqs.~(\ref{4.33}),
with the substitution: $L^2\rightarrow D(D-1)/|R|$.

The spin-2 result is quite expected in view of General Relativity. Einstein manifolds are nothing but the vacuum solutions of
Einstein equations. On such backgrounds, one can always consider linearized graviton fluctuations, which of course will
propagate consistently, thanks to General Relativity. Note that it is the absence of a stress-energy tensor that enables
one to take into account solely graviton fluctuations in the EoM's.

For $s=1$, no restrictions on the gravitational background are imposed. In this case, it is easy to see that
the gauge system will instead be described by:
\beq \left(\nb^2-R_{\m\n}u^\m d^\n\right)\F=0,\qquad \ddiv\F=0;\qquad\qquad \nb^2\l=0.\eeq{4.35}
In particular, the mass-like terms $M_0^2$ and $M_0^{\prime\,2}$ must be set to zero.

\subsection{Electromagnetic Background}\label{sec:gbem}

In this section, we will consider the propagation of a charged bosonic field in an EM background,
and will end up with a no-go for a higher-spin gauge boson, and a yes-go for a massless vector.
The EM counterparts of the involutive systems~(\ref{4.18})--(\ref{4.19}), for the spin-$s$
massless boson $\F$ and the accompanying spin-$(s-1)$ gauge parameter $\l$, read:
\beq \left[
           \begin{array}{cc}
             \dalb & 0 \\
             0 & \dalb' \\
           \end{array}
         \right]
\doub=0,\qquad \bar{g}_1\doub=0,\qquad \bar{g}_2\doub=0,\eeq{4.36}
with the deformed operators given directly from Eq.~(\ref{2.48}) as:
\begin{equation}\label{4.37} \begin{split}
\text{d'Alembertian}:~\dalb&=\mD^2-\bar{M}_0^2+2iqF_{\m\n}u^\m d^\n,\\
\text{Divergence}:~\bar{g}_1&=d\!\cdot\!\mD,\\
\text{Trace}:~\bar{g}_2&=d^2,\end{split}\end{equation}
along with $\dalb'=\dalb+\bar{M}_0^2-\bar{M}_0^{\prime\,2}$, where the mass-like terms $\bar{M}_0^2$ and
$\bar{M}_0^{\prime\,2}$ are polynomials in the index operator $N$ that vanish in the limit of zero background
EM field strength. On the other hand, the deformed symmetrized gradient is identified as a zeroth-order
deformation (for a reason analogous to that of the gravitational case), i.e.,
\beq \text{Symmetrized gradient}:~\bar{g}_{-1}=u\!\cdot\!\mD.\eeq{4.38}

In order to consider gauge transformations: $\d\F=\bar{g}_{-1}\l$, one needs the commutators of $\bar{g}_{-1}$
with the other operators; they are easy to compute. Upon using the Eqs.~(\ref{4.36}), one ends
up with the following variation of the involutive system:
\begin{equation}\label{4.39}\begin{split}
\d(\dalb\F)&~=~\left(\bar{g}_{-1}\bar{M}_0^{\prime\,2}-\bar{M}_0^2\bar{g}_{-1}\right)\l,\\
\d(\bar{g}_1\F)&~=~\left(\bar{M}_0^{\prime\,2}-3iqF_{\m\n}u^\m d^\n\right)\l,\\
\d(\bar{g}_2\F)&~=~0,
\end{split}\end{equation}
where the constancy of background field strength has been taken into account. It is clear that gauge invariance
cannot be restored for a generic spin $s>1$, irrespective of the mass parameters. In particular,
$\bar{M}_0^{\prime\,2}$ may only be a function of the index operator $N$, and so it cannot cancel the operation of
the $F_{\m\n}u^\m d^\n$-term in the variation $\d(\bar{g}_1\F)$.

Thus, we come up with a no-go theorem: a charged gauge boson with spin $s>1$ cannot propagate consistently
in an EM background. This agrees with the no-go result forbidding the minimal coupling of massless
higher-spin particles to a $U(1)$ gauge field~\cite{New}.

\subsubsection*{Yes-Go for Massless Vector}

For spin $s=1$, the right-hand sides of Eqs.~(\ref{4.39}) may all vanish, i.e., we have a yes-go result.
To see this, let us note that $d^\m\l=0$ in this case, and so the variation $\d(\bar{g}_1\F)$ vanishes if
$\bar{M}_0^{\prime\,2}$ is set to zero. Then, the variation $\d(\dalb\F)$ also vanishes with the choice
$\bar{M}_0^2=0$. This leaves us with the following involutive system for a massless vector $\F$:
\beq \left(\mD^2+2iqF_{\m\n}u^\m d^\n\right)\F=0,\qquad d\!\cdot\!\mD\,\F=0,\eeq{4.40}
in an EM background: $F_{\m\n}=\text{constant}$, along with the gauge symmetry:
\beq \d\F=u\!\cdot\!\mD\l,\qquad \mD^2\l=0.\eeq{4.41}

This yes-go result may not come as a surprise given the existence of Yang-Mills theories as consistent
interacting theories of spin-1 gauge fields. Indeed, the system~(\ref{4.40})--(\ref{4.41}) can be obtained
from a non-Abelian gauge theory linearized around some background. To see this, let us consider an $SU(2)$ gauge
field $W_\m^a$, whose field strength is given by: $G_{\m\n}^a=\de_\m W_\n^a-\de_\n W_\m^a+g\e^{abc}W_\m^b W_\n^c$,
where $g$ is the Yang-Mills coupling. The EoM's are:
\beq \de^\m G_{\m\n}^a+g\e^{abc}W^{\m,b}G_{\m\n}^c=0,\qquad a=1,2,3,\eeq{4.42}
and the infinitesimal gauge transformations read:
\beq \d W_\m^a=\de_\m\l^a+g\e^{abc}W^{\m,b}\l^c.\eeq{4.43}
It is easy to see that the EoM's~(\ref{4.42}) admit the following solution:
\beq W_\m^1=W_\m^2=0,\qquad W_\m^3=A_\m\neq0,\quad\text{with}\quad F_{\m\n}=2\de_{[\m} A_{\n]}
=\text{constant}.\eeq{4.44}

On this background, let us now consider small fluctuations $w_\m^a$. At the linearized level,
the mode $w_\m^3$ behaves as if it were a $U(1)$ gauge field:
\beq \de^\m\left(\de_\m w_\n^3-\de_\n w_\m^3\right)=0,\qquad \d w_\m^3=\de_\m\l^3.\eeq{4.45}
The other two modes have the linearized field strengths:
\beq f_{\m\n}^i\equiv2\left(\de_{[\m} w_{\n]}^i+(-)^igA_{[\m}w^{j\neq i}_{\n]}\right),
\qquad i,j=1,2,\eeq{4.46}
through which these modes are described by the coupled equations:
\beq \de^\m f_{\m\n}^i+(-)^ig\left(A^\m f_{\m\n}^{j\neq i}-F_{\m\n}w^{\m,j\neq i}\right)=0,\eeq{4.47}
that are invariant under the gauge transformations:
\beq \d w_\m^i=\de_\m\l^i+(-)^igA_\m\l^{j\neq i}.\eeq{4.48}

Now, we consider the following complex vector field and gauge parameter:
\beq \F_\m\equiv\tfrac{1}{\sqrt{2}}\left(w_\m^1+iw_\m^2\right),\qquad
\l\equiv\tfrac{1}{\sqrt{2}}\left(\l^1+i\l^2\right).\eeq{4.49}
At the linear level, the Yang-Mills coupling $g$ can now be identified as the $U(1)$ charge $q$ of the
vector $\F_\m$, on which acts the covariant derivative: $\mD_\m\equiv\de_\m+igA_\m$. The EoM's
and the gauge symmetry of $\F_\m$ read:
\beq 2\mD^\m\mD_{[\m}\F_{\n]}-iqF_{\m\n}\F^\m=0,\qquad \d\F_\m=\mD_\m\l,\eeq{4.50}
which reduces precisely to our system~(\ref{4.40})--(\ref{4.41}) in the Lorenz gauge: $\mD^\m\F_\m=0$.

\section{Massless Fermionic Fields}\label{sec:gf}

This section explores the massless limit of the involutive system of a massive higher-spin fermion. In this limit, as we will see, the flat-space involutive system~(\ref{3.5}) acquires a gauge symmetry with an ``unfree'' gauge parameter
governed by the same involutive system as the field. This is nothing but the description of a massless higher-spin fermion in the transverse-traceless gauge.
We confirm, along the line of Appendix~\ref{sec:constgauge}, that the resulting involutive system describes the correct number of physical DoF's of a gauge
fermion. Given this reformulation, we go on to studying the consistent free propagation of higher-spin gauge fermions in nontrivial backgrounds.

\subsection{Minkowski Background}\label{sec:gfflat}

Let us consider, for the massive rank-$n$ fermion $\Ps$ of Eqs.~(\ref{3.5}),
the transformation:
\beq \d\Ps=g_{-1}\ve,\qquad \ve=\tfrac{1}{(n-1)!}\,\ve_{\m_1\cdots\m_{n-1}}(x)u^{\m_1}\cdots u^{\m_{n-1}},\eeq{5.1}
where $g_{-1}$ is the symmetrized gradient operator, already introduced in
Eq.~(\ref{4.2}). In view of the commutation relations for the symmetrized gradient:
\beq [f_{0},g_{-1}]=0,\qquad[g_{1},g_{-1}]=(f_{0}+m)^2,\qquad[f_{1},g_{-1}]=f_{0}+m,\eeq{5.2}
it is easy to see that the left-hand sides of Eqs.~(\ref{3.5}) transform as:
\bea \d(f_0\Ps)&=&g_{-1}(f_0\ve),\nonumber\\
\d(g_1\Ps)&=&g_{-1}(g_1\ve)+(f_{0}+m)^2\ve,\label{5.3}\\
\d(f_1\Ps)&=&g_{-1}(f_1\ve)+(f_{0}+m)\ve.\nonumber\eea{ekxsnsl5}

To see if transformations of the type~(\ref{5.1}) may become a symmetry
of the Fierz-Pauli system~(\ref{3.5}), let us require that
$\ve$ itself be governed by the following involutive equations:
\beq f_0\ve=0,\qquad g_1\ve=0,\qquad f_1\ve=0.\eeq{5.4}
Then, then the variations~(\ref{5.4}) vanish if and only if:
\beq m\ve=0.\eeq{5.5}
Clearly, in the zero mass limit: $m\rightarrow0$, there appears a nontrivial gauge symmetry.

The involutive system of a massless fermion therefore enjoys a gauge symmetry~(\ref{5.1}), where the gauge parameter itself is governed by Eqs.~(\ref{5.4}) with zero mass. In this case, the massless Dirac
operator $\dir$ is of relevance, for which we have the following:
\beq \text{Massless Dirac:}\quad \dir~\equiv~\des=\lim_{m\rightarrow0}f_0,\qquad \text{with}\quad [N,\dir]=0.
\eeq{5.6}
Note that the set of operators essential for the massless case is given by:
\beq \underline{\mathcal{F}}=\{\dir,g_1,f_1,g_{-1}\},\eeq{5.7}
and that the massless counterparts of the commutators~(\ref{5.2}) read:
\beq [\dir,g_{-1}]=0,\qquad[g_{1},g_{-1}]=\dir^2,\qquad[f_{1},g_{-1}]=\dir.\eeq{5.8}
These commutators close completely within the set $\underline{\mathcal{F}}\backslash\left\{g_{-1}\right\}$.
This ensures transverse-traceless gauge symmetry, which in turn controls the DoF count, as we will now show.

Let us recall that the rank-$n$ gauge field $\Ps$ and the rank-$(n-1)$ gauge parameter $\ve$ are both governed
by the same involutive set of equations:
\beq \dir\douf=0,\qquad g_1\douf=0,\qquad f_1\douf=0.\eeq{5.9}
It is easy to see from Section~\ref{sec:mfflat} that the massless limit does not affect the involutive structure
of the Fierz-Pauli system~(\ref{3.5}). Neither does it alter the DoF count~(\ref{3.12}). However, because of the
emergence of (unfree) gauge symmetry, the count~(\ref{3.12}) includes pure gauge modes as well. In this case, the
analysis of Appendix~\ref{sec:constgauge} applies, and one can easily write down the number of physical DoF for a
rank-$n$ gauge fermion. As seen from formula~(\ref{A24}), it must be the difference between the DoF count of a
massive rank-$n$ fermion and that of a massive rank-$(n-1)$ fermion:
\beq \mathfrak{D}_f^{(0)}(n)=\mathfrak{D}_f(n)-\mathfrak{D}_f(n-1).\eeq{5.10}
From the DoF count formula~(\ref{3.12}), then it follows that
\beq \mathfrak{D}_f^{(0)}(n)={D+n-4\choose n}\,2^{[D-2]/2}.\eeq{5.11}
This is indeed the correct number of physical DoF's for a rank-$n$ gauge fermion~\cite{Rahman:2015pzl}.

\subsection{Gravitational Background}\label{sec:gfgrav}

We would like to have the deformed counterparts of the operators~(\ref{5.7})
in a gravitational background; they constitute the following set:
\beq \underline{\hat{\mathcal{F}}}=\{\dirh,\hat{g}_1,\hat{f}_1,\hat{g}_{-1}\},\eeq{5.12}
which includes the operators appearing in the zero mass limits of Eqs.~(\ref{3.26}), i.e.,
\beq \dirh=\nbs-\mm_0,\qquad \hat{g}_1=\ddiv,\qquad\hat{f}_1=\ds,\eeq{5.13}
where, as we recall from the ansatz~(\ref{3.16}), the mass-like term takes the form:
\beq \mm_0=\m\left(N+\d\right),\eeq{5.14}
with $\m$ being a mass parameter that vanishes in the zero-curvature limit, and $\d$ a
numerical constant. In order to write down the deformed symmetrized gradient $\hat{g}_{-1}$, we recall
that Eq.~(\ref{3.24}) sets to infinity the suppression scale $\L$ of the higher-dimensional operators.
This leaves us with the following generic form of $\hat{g}_{-1}$:
\beq \hat{g}_{-1}=\dgrad-\hat{\m}\,\us,\eeq{5.15}
where $\hat{\m}$ is another constant mass parameter vanishing in the limit of zero curvature. Note that
Eq.~(\ref{5.15}) is in contrast with its bosonic counterpart~(\ref{4.17}), where only the zeroth-order
deformation could be written down. The higher-spin gauge fermion $\Ps$ and the gauge parameter $\ve$ are
governed by the following involutive systems:
\beq \left[
           \begin{array}{cc}
             \dirh & 0 \\
             0 & \dirh^{\,\prime} \\
           \end{array}
         \right]
\douf=0,\qquad \hat{g}_1\douf=0,\qquad \hat{f}_1\douf=0,\eeq{5.16}
where $\dirh^{\,\prime}=\dirh+\mm_0-\mm_0^\prime$, for some mass-like term $\mm_0^\prime$
of the form~(\ref{5.14}).

We consider gauge transformations of the gauge-fermion involutive system: $\d\Ps=\hat{g}_{-1}\ve$.
In view of Eqs.~(\ref{5.16}), it is easy to obtain the following variations:
\begin{equation}\label{5.17}\begin{split}
\d(\dirh\Ps)&~=~[\dirh,\hat{g}_{-1}]\ve+\hat{g}_{-1}(\mm_0'-\mm_0)\ve,\\
\d(\hat{g}_1\Ps)&~=~[\hat{g}_1,\hat{g}_{-1}]\ve,\\
\d(\hat{f}_1\Ps)&~=~[\hat{f}_1,\hat{g}_{-1}]\ve.
\end{split}\end{equation}
In order see how these variations may vanish, we need the commutators of $\hat{g}_{-1}$
with the other operators: $\{\dirh,\hat{g}_1,\hat{f}_1\}$. The simplest one reads:
\beq [\hat{f}_1,\hat{g}_{-1}]=\dirh^{\,\prime}+2\hat{\m}\,\us\hat{f}_1+\mm_0'-2\hat{\m}\left(N+D/2\right).
\eeq{5.18}
The vanishing of the variation $\d(\hat{f}_1\Ps)$ therefore requires that
\beq \mm_0'=2\hat{\m}\left(N+D/2\right).\eeq{5.19}
Next, the computation of $[\dirh,\hat{g}_{-1}]$ is simplified by noting that, in the limit of $m\rightarrow0$ and
$\L\rightarrow\infty$, the hermitian conjugate (in the sense of footnote~5) of Eq.~(\ref{3.17}) provides with
$[\dirh,\dgrad]$, whereas the commutator $[\dirh,\us]$ is easy to compute. The end result is:
\beq [\dirh,\hat{g}_{-1}]=2\hat{\m}\,\us\dirh^{\,\prime}+\rad u^2\hat{f}_1-\left(\m+2\hat{\m}\right)
\hat{g}_{-1}+\us\left[2\hat{\m}\left(\mm_0'-\hat{\m}\right)-\rad\left(N-\tfrac{1}{2}+D/2\right)\right],
\eeq{5.20}
where we used the maximal symmetry of the background. When plugged into the variation
$\d(\dirh\Ps)$, the last term of Eq.~(\ref{5.20})$-$combined with the result~(\ref{5.19})$-$implies:
\beq \hat{\m}^2=\tfrac{1}{4L^2}.\eeq{5.21}
The terms containing $\hat{g}_{-1}$, on the other hand, justify the mass ansatz~(\ref{5.14}), and give:
\beq \mm_0=2\hat{\m}\left(N-2+D/2\right).\eeq{5.22}
This completely fixes all the parameters in the theory. It is conventional to choose the positive root of
Eq.~(\ref{5.21})~\cite{Metsaev:2006zy,Metsaev:2013wza}, which sets: $\hat{\m}=+\tfrac{1}{2L}$.

One still needs to show that the variation $\d(\hat{g}_1\Ps)$ also vanishes. Given the relation~(\ref{C4}),
it is straightforward to cast the commutator $[\hat{g}_1,\hat{g}_{-1}]$ into the following form:
\beq [\hat{g}_1,\hat{g}_{-1}]=\nb^2-\hat{\m}(\dirh^{\,\prime}+\mm_0')
+\tfrac{1}{L^2}\left[u^2\hat{f}_1^2+\tfrac{1}{2}\us\hat{f}_1-N\left(N+D-\tfrac{3}{2}\right)\right].\eeq{5.23}
The expression of $\nb^2$ in terms of the massless Dirac operator is somewhat subtle. One needs to compute the
anti-commutator $\{\dirh^{\,\prime},\dirh^{\,\prime}\}$ to show that:
\beq \nb^2=\dirh^{\,\prime2}+2\mm_0'\dirh^{\,\prime}+\mm_0^{\prime\,2}+\tfrac{1}{L^2}
\left[\us\hat{f}_1-N-\tfrac{1}{4}D(D-1)\right].\eeq{5.24}
Then, the expressions~(\ref{5.23})--(\ref{5.24}) indeed renders the variation $\d(\hat{g}_1\Ps)$ vanishing
on account of the relations~(\ref{5.19}) and~(\ref{5.21}).

We are now in a position of presenting our gauge invariant involutive system. The rank-$n$ gauge field $\Ps$
and the rank-$(n-1)$ gauge parameter $\ve$ obey:
\begin{equation}\label{5.25}\begin{split}
\left(\nbs-\mm_0\right)\Ps&=0,\qquad \ddiv\Ps=0,\qquad \ds\,\Ps=0,\\
\left(\nbs-\mm_0'\right)\ve&=0,\qquad \ddiv\ve=0,\qquad \ds\,\ve=0,
\end{split}\end{equation}
where the mass-like terms are given by:
\beq \mm_0L=N-2+D/2,\qquad \mm_0'L=N+D/2,\eeq{5.26}
and the gauge transformations read:
\beq \d\Ps=\left(\dgrad-\tfrac{1}{2L}\us\right)\ve.\eeq{5.27}
This system holds good for an arbitrary-spin gauge fermion in $D\geq3$ only in maximally symmetric spaces. While
the length scale $L$ appearing in Eqs.~(\ref{5.26})--(\ref{5.27}) is an AdS radius, the analytic continuation $L\rightarrow iL$
will aptly describe a Dirac fermion in dS space.

\subsubsection*{Rarita-Schwinger Gauge Field}

For $s=\tfrac{3}{2}$, the gravitational background will have a weaker constraint, but the
involutive system~(\ref{5.25})--(\ref{5.27}) holds good, with $L\rightarrow\sqrt{D(D-1)/|R|}$.
Let us recall from Section~\ref{sec:mfgrav} that the massive
involutive system is consistent in Einstein spaces. Going massless in this case, by requiring gauge symmetry, does not
pose any additional condition. To see this, let us notice how the gauge variations~(\ref{5.17}) could vanish for generic spin.
The conditions on the background played r\^ole only through Eqs.~(\ref{5.20}),~(\ref{5.23}) and~(\ref{5.24}). An Einstein
manifold may well be conformally non-flat, i.e., possess a non-vanishing Weyl tensor. In this case, the right-hand
side of Eq.~(\ref{5.20}) picks up an additional term: $W_{\m\n\r\s}\g^\m u^\n u^\r d^\s$, which
gives zero contribution in the variation $\d(\dirh\Ps)$, since $d^\m\ve=0$. Similarly, Eq.~(\ref{5.23})
would include terms containing a Weyl tensor and at least one $d^\m$, and they do not contribute to the variation
$\d(\hat{g}_1\Ps)$. Last but not the least, Eq.~(\ref{5.24}) also picks up the term: $W_{\m\n\r\s}\g^{\m\n}\g^{\r\s}$.
By using the symmetries of the Weyl tensor, the $\g$-matrix product can be rewritten as:
$\g^{\m\n\r\s}-2g^{\m\r}g^{\n\r}$. The latter terms give zero on account of the Bianchi identity and tracelessness
of the Weyl tensor. Therefore, it is necessary and sufficient to require that the background be an Einstein space.

This result makes sense from the perspective of supergravity. The classical solutions of pure $\mathcal{N}=1$
supergravity are indeed Einstein spaces, on which fluctuations of the massless spin-$\tfrac{3}{2}$ Majorana fermion
propagate consistently. However, extended supergravity theories admit more generic classical backgrounds. In particular,
Maxwell-Einstein spaces appear in pure $\mathcal{N}=2$ (un)gauged supergravity, and this seems to contradict our result.
One of the loopholes lies in the deformed gauge transformation of the gravitino; it involves a $U(1)$ gauge
field~\cite{Ferrara:1976fu,Freedman:1976aw,Boulanger:2018fei}$-$a possibility we do not consider. Moreover, in
the gauged theory the complex gravitino has a $U(1)$ charge as well.

\subsection{Electromagnetic Background}\label{sec:gfem}

Let us recall that in Section~\ref{sec:mfem} we assumed minimal coupling, i.e.,
a nonzero charge $q$ of the higher-spin fermion. However, it is manifest that the resulting involutive system~(\ref{3.44})--(\ref{3.45}) is ill-defined in the massless limit: $m\rightarrow0$. This can be traced back to Eqs.~(\ref{lin}), which admit no solutions of the deformed Dirac, divergence and $\g$-trace operators as the mass goes to zero for spin $s\geq\tfrac{3}{2}$. Thus, we are lead to a no-go theorem: a charged gauge fermion cannot propagate consistently in a purely EM background.
In other words, there is no consistent theory of a gauge fermion, minimally coupled to a $U(1)$ field, that admits a pure background of the Maxwell field as a classical solution. This is in accordance with the no-go results~\cite{New,Henneaux:2012wg} that forbid in flat space the minimal coupling of a massless fermion with spin $s\geq\tfrac{3}{2}$ to a $U(1)$ gauge field.

One way to bypass this no-go is to consider additional
interactions in the theory such that purely $U(1)$ backgrounds are not allowed. This works at least for a massless charged Rarita-Schwinger field, which requires a cosmological constant~\cite{Zachos:1978iw} (see also~\cite{Boulanger:2018fei} for a cohomological derivation). Indeed, $\mathcal{N}=2$ gauged supergravity~\cite{Ferrara:1976fu,Freedman:1976aw} consistently incorporates a massless gravitino minimally coupled to a $U(1)$ field (graviphoton) as well as gravity in the presence of a cosmological constant. Determined by Eq.~(\ref{5.26}),
the mass parameter in this case is also related to the $U(1)$ charge. In $\text{AdS}_4$ the relations read:
\beq \mm_0^2=1/L^2=2q^2M_P^2.\eeq{5.28}
The classical solutions of pure $\mathcal{N}=2$ are, of course, Maxwell-Einstein spaces on which fluctuations of the massless charged
gravitino propagate consistently. Whether a similar type of yes-go can be found for higher-spin gauge fermions is an open question.

\section{Partially Massless Fields}\label{sec:PMAdS}

In a constant curvature space, it turns out that gauge symmetries of a higher-spin field appear for a discrete series of mass parameters,
known as partially massless (PM) points. Originally studied in~\cite{Deser:1983mm,Higuchi:1986py}, this phenomenon
was further investigated in~\cite{Deser:2001pe,Zinoviev:2001dt,Skvortsov:2006at,Francia:2008hd,Kuzenko:2019kqw}.
In this section, we consider the involutive system of PM bosons and fermions.
Just like a massless system is described by Eqs.~(\ref{4.33}) or~(\ref{5.25}),
a PM field and its gauge parameter are also governed by the same type of involutive systems. However, PM fields are more general in that their gauge transformations may include multiple gradients of the gauge parameters.
A PM field is said to have depth $(k+1)$ when its gauge transformation contains $(k+1)$ space-time derivatives plus possibly a lower-derivative tail:
\begin{equation}\label{6.1}\begin{split}
\text{Boson}:~\d\Fpm&=\left[(\dgrad)^{k+1}+\cdots\right]\lpm,\qquad k=0,1,\cdots,s-1,\\
\text{Fermion}:~\d\Pspm&=\left[(\dgrad)^{k+1}+\cdots\right]\vepm,\qquad k=0,1,\cdots,n-1,
\end{split}\end{equation}
where the subscripts on the fields and gauge parameters denote their respective ranks (unlike that on an operator, which gives the
negative of its weight), whereas the superscript on a PM field
stands for its depth. Let us denote by
$\hat{g}_{-k-1}$ the weight-$(k+1)$ operators appearing in the PM gauge transformations~(\ref{6.1}):
\beq \hat{g}_{-k-1}=(\dgrad)^{k+1}+\text{lower-derivative tail}.\eeq{6.2}
Note that the strictly massless case corresponds to $\text{depth}=1$, i.e., $k=0$.
We would like to find the explicit form of $\hat{g}_{-k-1}$, i.e., that of the depth-$(k+1)$ gauge transformations~(\ref{6.1})
as well as the PM discrete points of the mass parameters\footnote{With some abuse of notations, we will denote the discrete mass points by $\mm_k$ and $\mm_k^{\prime}$ respectively for the PM field and the gauge parameter.
These mass parameters are of course $w=0$ operators, for which the subscript
$k$ does \emph{not} correspond to the weight but to the value of depth minus one. Accordingly, the mass parameters
in the strictly massless case are denoted by $\mm_0$ and $\mm_0^\prime$, as in Eqs.~(\ref{4.32}) or~(\ref{5.26}).} in AdS space.

The DoF count works in the following way. As we will see, just like the strictly massless case, the PM field and its gauge parameter will both be governed by
their respective involutive systems. Therefore, the analysis of Appendix~\ref{sec:constgauge} also applies here; the number of physical DoF will simply be the difference between the DoF counts of a massive field and a massive gauge parameter:
\begin{equation}\label{6.4}\begin{split}
\text{Boson}:~\mathfrak{D}_b^{(k)}(s)&=\mathfrak{D}_b(s)-\mathfrak{D}_b(s-k-1),\\
\text{Fermion}:~\mathfrak{D}_f^{(k)}(n)&=\mathfrak{D}_f(n)-\mathfrak{D}_f(n-k-1).
\end{split}\end{equation}
Then, the DoF count at depth $(k+1)$ follows directly from formula~(\ref{2.12}) or~(\ref{3.12}).
Below we go into the details separately for bosonic and fermionic PM fields.

\subsection{Bosonic Fields}\label{sec:PMb}

For bosonic PM fields, it will be convenient to define the following operator:
\beq\Box\equiv[\ddiv,\dgrad],\eeq{6.3}
which can be written in terms of $\nb^2$ through Eq.~(\ref{C4}) in AdS space.
In analogy with the strictly massless case of Section~\ref{sec:gbgrav}, the involutive system of a
spin-$s$ depth-$(k+1)$ PM boson $\Fpm$ and its gauge parameter $\lpm$ can be written as:
\beq \left[
           \begin{array}{cc}
             \hat{g}_0^{(k+1)} & 0 \\
             0 & \hat{g}_0^{\prime(k+1)} \\
           \end{array}
         \right]
\doupmb=0,\qquad \hat{g}_1\doupmb=0,\qquad \hat{g}_2\doupmb=0,\eeq{6.5}
where $\hat{g}_1=\ddiv$ and $\hat{g}_2=d^2$ are the usual divergence and trace operators appearing in Eq.~(\ref{4.16}),
while the deformed d'Alembertian operators $\hat{g}_0^{(k+1)}$ and $\hat{g}_0^{\prime(k+1)}$ generalize Eqs.~(\ref{4.32})
for arbitrary depth. We will prove that the d'Alembertians are given by:
\begin{equation}\label{6.6}\begin{split}
\hat{g}_0^{(k+1)}&~=~\Box-\rad(k+2)(k-2N-D+3)~=~\nb^2-\mm_k^2+\rad u^2 d^2,\\
\hat{g}_0^{\prime(k+1)}&~=~\Box-\rad k(k+2N+D-1)~=~\nb^2-\mm_k^{\prime\,2}+\rad u^2 d^2,
\end{split}\end{equation}
where the PM mass parameters at depth $(k+1)$ are specified as:
\begin{equation}\label{6.7}\begin{split}
\mm_k^2L^2&\equiv(N-k-2)(N-k+D-3)-N,\\
\mm_k^{\prime\,2}L^2&\equiv (N+k)(N+k+D-1)-N.
\end{split}\end{equation}
We will also prove the following explicit form of the gauge transformations:
\beq \hat{g}_{-k-1}=\left(\dgrad\right)^{\e_k}\left[\left(\dgrad\right)^2-\rad u^2(N-s+1)^2\right]^{[k+1]/2},\eeq{6.8}
where $\e_k=\tfrac{1}{2}\left[1+(-)^k\right]$, which is 1(0) for $k$ even(odd). Note that Eq.~(\ref{6.8}) induces the
the following iterative expression on a spin-$(s-k-1)$ gauge parameter:
\beq \hat{g}_{-k-1}=\left[\left(\dgrad\right)^2-\rad c_ku^2\right]\hat{g}_{-k+1},\qquad c_k=k^2,\qquad k\geq2.\eeq{6.9}

In what follows we provide a proof of Eqs.~(\ref{6.5})--(\ref{6.9}) by recourse to the method of induction.
To proceed, let us make the following ans\"atze for the deformed d'Alembertians:
\beq \hat{g}_0^{(k+1)}=\Box-\rad\left(a_kN+b_k\right),\qquad
\hat{g}_0^{\prime(k+1)}=\Box-\rad\left(a'_kN+b'_k\right),\eeq{6.10}
where $a_k, b_k$ are their primed counterparts are numerical constants. Therefore, in order to prove
Eqs.~(\ref{6.6})--(\ref{6.7}) we ought to show the following:
\beq a_k=-2(k+2),\quad b_k=(k+2)(k-D+3),\quad a'_k=2k,\quad b'_k=k(k+D-1).\eeq{6.11}
Similarly, the PM gauge transformations will also be proved with ans\"atze
compatible with Eqs.~(\ref{6.8})--(\ref{6.9}).
Below present our proofs for $k=0,1,2$, and then for generic $k$.
\newpage
\noindent\underline{$k=0$}: This is the strictly massless case, for which Eqs.~(\ref{6.5})--(\ref{6.7}) have already been
proved in Section~\ref{sec:gbgrav}. Indeed, for $k=0$ the gauge transformation is given by: $\hat{g}_{-1}=\dgrad$,
whereas the dynamical equations reduce to Eqs.~(\ref{4.32}) given the trace constraints.\newline\vspace{-10pt}\newline
\noindent\underline{$k=1$}: This corresponds to depth 2$-$the simplest nontrivial PM gauge symmetry. In this case, the most
generic form of the PM gauge transformation could be:
\beq \hat{g}_{-2}=\left(\dgrad\right)^2-\rad c_1u^2,\qquad c_1=\text{constant}.\eeq{6.12}
In order to compute the gauge variations of the EoM's we need the commutator of $\hat{g}_{-2}$ with
$\{\hat{g}_0^{(2)},\hat{g}_1,\hat{g}_2\}$, which are given in Eqs.~(\ref{C11})--(\ref{C13}). Upon making use of the
involutive system for the gauge parameter $\l_{s-2}$, these variations simplify to Eqs.~(\ref{C14})--(\ref{C15}).
Consequently, gauge invariance requires the following choice of constants:
\beq a_1=-6,\qquad b_1=-3(D-4),\qquad a'_1=2,\qquad b'_1=D,\qquad c_1=1.\eeq{6.13}
These are precisely the values given for $k=1$ by Eqs.~(\ref{6.11}) and the gauge transformation~(\ref{6.8}),
with $c_1$ being the eigenvalue of $(N-s+1)^2$ corresponding to $\l_{s-2}$.
\newline\vspace{-10pt}\newline
\noindent\underline{$k=2$}: Let us make the ansatz that the depth-$3$ gauge transformation is implemented by:
\beq \hat{g}_{-3}=\dgrad\left[\left(\dgrad\right)^2-\rad c_2u^2\right],
\qquad c_2=\text{constant}.\eeq{6.14}
The variations of the EoM's of the PM field $\F^{(3)}_s$ is easy to compute given the basic commutation relations~(\ref{C10})--(\ref{C13}). Again, the involutive system for the gauge parameter $\l_{s-3}$ is taken into account in order to simplify these gauge variations. Their
explicit forms are given in Eqs.~(\ref{C16})--(\ref{C17}). In order for the gauge variations to vanish we must have:
\beq a_2=-8,\qquad b_2=-4(D-5),\qquad a'_2=4,\qquad b'_2=2(D+1),\qquad c_2=4.\eeq{6.15}
Again, these are the values Eqs.~(\ref{6.11}) and the gauge transformation~(\ref{6.8}) give for $k=2$.
Here, $c_2$ is indeed the eigenvalue of $(N-s+1)^2$ corresponding to $\l_{s-3}$. Also, the recursion formula~(\ref{6.9})
works, since setting $k=2$ therein reproduces Eq.~(\ref{6.14}) with $c_2=4$.
\newline\vspace{-10pt}\newline
\noindent\underline{Generic $k$}: Let us assume that the involutive system~(\ref{6.5})--(\ref{6.9}) holds good up to and
including $k=j-2$, for some integer $j\geq2$. It will then follow that the same system also consistently describes the case
$k=j$. To see this, let us make the ansatz that the depth-$(j+1)$ PM gauge transformation
is implemented by the following operator:
\beq \hat{g}_{-j-1}=\left[\left(\dgrad\right)^2-\rad c_j u^2\right]\hat{g}_{-j+1},\eeq{6.16}
where $c_{j}$ is some constant to be determined. Recall that for the deformed d'Alembertians we have the ans\"atze~(\ref{6.10}).
Then we can compute the gauge variations of the left-hand sides of the involutive equations for $\F_s^{(j+1)}$. They take the
following form:
\bea
&\d\!\left[\hat{g}_0^{(j+1)}\F_s^{(j+1)}\right]=\left[\Box-\rad\left(a_j N+b_j\right)\right]
\left[\left(\dgrad\right)^2-\rad c_j u^2\right]\hat{g}_{-j+1}\l_{s-j-1},&\nonumber\\
&\d\!\left[\hat{g}_1\F_s^{(j+1)}\right]=\ddiv\left[\left(\dgrad\right)^2-\rad c_j u^2\right]\hat{g}_{-j+1}\l_{s-j-1},&\label{6.17}\\
&\d\!\left[\hat{g}_2\F_s^{(j+1)}\right]=d^2\left[\left(\dgrad\right)^2-\rad c_j u^2\right]\hat{g}_{-j+1}\l_{s-j-1},&\nonumber
\eea{C177777}
with the ``unfree'' gauge parameter $\l_{s-j-1}$ being subject to:
\beq \left[\Box-\rad\left(a'_jN+b'_j\right)\right]\l_{s-j-1}=0,\qquad \ddiv\l_{s-j-1}=0,\qquad d^2\l_{s-j-1}=0,\eeq{6.18}
where $a_{j}$ and $b_{j}$ and their primed counterparts are constants to be determined.

In computing the right-hand sides of Eqs.~(\ref{6.17}), one needs to make repeated use of the commutators~(\ref{C10})--(\ref{C13}),
and conditions~(\ref{6.18}) on the gauge parameter. After a tedious but straightforward calculation, one arrives at the following results:
\bea
\d\!\left[\hat{g}_0^{(j+1)}\F_s^{(j+1)}\right]&=&\left\{\rad\left(\dgrad\right)^{j+1}\mathcal{L}_{j1}+\cdots\right\}\l_{s-j-1},
\nonumber\\
\d\!\left[\hat{g}_1\F_s^{(j+1)}\right]&=&\left\{\rad\left(\dgrad\right)^{j}\mathcal{L}_{j2}+\cdots\right\}\l_{s-j-1},\label{6.19}\\
\d\!\left[\hat{g}_2\F_s^{(j+1)}\right]&=&\left\{\rad\left(\dgrad\right)^{j-1}\mathcal{L}_{j3}+\cdots\right\}\l_{s-j-1},
\nonumber\eea{C199999}
where the ellipses stand for lower-derivative terms, and the $\mathcal{L}_j$'s are given by:
\begin{equation}\label{6.20}\begin{split}
\mathcal{L}_{j1}&=[a'_j-a_j-4(j+1)]N+[b_j'-b_j-(j+1)(a_j+2(j+D-1))],\\
\mathcal{L}_{j2}&=(j+1)(a'_j-2j)N\\
&~~~+[(j+1)(b_j'-D+1)-2c_j-(j-1)(j(j+D-1)+D-1)],\\
\mathcal{L}_{j3}&=[j(j+1)a'_j-4c_j-2j^2(j-1)]N\\
&~~~+[j(j+1)b'_j-2(D+2(j-1))c_j-j^2(j-1)(j+D-3)].
\end{split}\end{equation}
In deriving the above expressions one makes use of the assumption that the involutive system~(\ref{6.5})--(\ref{6.8}) holds
good for $k\leq j-2$. Thus, the expressions~(\ref{6.9})--(\ref{6.11}) are valid
up to and including $k=j-2$. Now, in order for the gauge variations~(\ref{6.19}) to vanish it is necessary that the gauge parameter $\l_{s-j-1}$ belongs simultaneously to the kernels of $\mathcal{L}_{j1}$, $\mathcal{L}_{j2}$ and $\mathcal{L}_{j3}$.
It is however easy to see that, for a nontrivial gauge parameter, such conditions can only be satisfied when the operators themselves vanish. This gives a unique
set of solutions for $c_j, a_j, b_j, a'_j$ and $b'_j$; it coincides with that given by Eqs.~(\ref{6.9}) and~(\ref{6.11}) for $k=j$. Too see that these values also suffice for the vanishing of the gauge variations~(\ref{6.19}), one needs to compute all the lower-derivative terms omitted in the ellipses. While one can convince oneself by explicitly working them out for any given $j$, we choose not to present this tedious exercise, and conclude without further ado.

Let us now summarize the results. In AdS space, the involutive system of a spin-$s$ depth-$(k+1)$ PM boson $\Fpm$ and its spin-$(s-k-1)$ gauge parameter $\lpm$ reads:
\begin{equation}\label{6.21}\begin{split}
\left(\nb^2-\mathfrak{m}_k^2\right)\Fpm&=0,\qquad \ddiv\Fpm=0,\qquad d^2\Fpm=0,\\
\left(\nb^2-\mathfrak{m}_k^{\prime\,2}\right)\lpm&=0,\qquad \ddiv\lpm=0,\qquad d^2\lpm=0,
\end{split}\end{equation}
with the mass terms given by Eqs.~(\ref{6.7}) for $k\geq0$. Note that the
above system has been presented without the $u^2d^2$-terms appearing in the d'Alembertians~(\ref{6.6}). This is possible
because the trace conditions themselves are a part of the involutive system. The depth-$(k+1)$ PM gauge symmetry transformations
of the system~(\ref{6.21}) are of the form:
\beq \d\Fpm=\hat{g}_{-k-1}\lpm,\eeq{6.22}
where the operator $\hat{g}_{-k-1}$ contains up to $(k+1)$ derivatives, given explicitly in Eq.~(\ref{6.8}).

\subsection{Fermionic Fields}\label{sec:PMf}

For fermionic PM fields, let us define a deformed covariant derivative $\D_\m$
as follows:
\beq \D_\m\equiv\nb_\m-\tfrac{1}{2L}\g_\m,\qquad [\D_\m,\D_\n]=-\rad\left(2u_{[\m}d_{\n]}\right).\eeq{6.23}
In analogy with the strictly massless case of Section~\ref{sec:gfgrav}, the involutive system of a
rank-$n$ depth-$(k+1)$ PM fermion $\Pspm$ and its gauge parameter $\vepm$ can be written as:
\beq \left[
           \begin{array}{cc}
             \hat{f}_0^{(k+1)} & 0 \\
             0 & \hat{f}_0^{\,\prime\,(k+1)} \\
           \end{array}
         \right]
\doupmf=0,\qquad \hat{g}_1^{\,\prime}\doupmf=0,\qquad \hat{f}_1\doupmf=0,\eeq{6.24}
where $\hat{f}_0^{(k+1)}$ and $\hat{f}_0^{\,\prime\,(k+1)}$ are
the deformed Dirac operators, while $\hat{g}_1^{\,\prime}\equiv\Ddiv=\ddiv-\tfrac{1}{2L}\ds$ is a deformed divergence,
and $\hat{f}_1$ the usual $\g$-trace operator.
We will show that:
\begin{equation}\label{6.25}\begin{split}
\hat{f}_0^{(k+1)}&~=~\Ds-\radf(N-k-2)~=~\nbs-\mm_k,\\
\hat{f}_0^{\,\prime\,(k+1)}&~=~\Ds-\radf(N+k)~=~\nbs-\mm_k^\prime,
\end{split}\end{equation}
where the PM mass parameters at depth $(k+1)$ generalize Eqs.~(5.26),
and are given by:
\beq \mm_k L\equiv N-k-2+D/2,\qquad \mm_k^\prime L\equiv N+k+D/2.\eeq{6.26}
The gauge transformations will be quite similar to the bosonic ones~(\ref{6.8}). Explicitly,
\beq \hat{g}_{-k-1}=\left(\Dgrad\right)^{\e_k}\left[\left(\Dgrad\right)^2-\rad u^2(N-n+1)^2\right]^{[k+1]/2}.\eeq{6.27}
Again, this induces the following iterative expression on a rank-$(n-k-1)$ gauge parameter:
\beq \hat{g}_{-k-1}=\left[\left(\Dgrad\right)^2-\rad \d_ku^2\right]\hat{g}_{-k+1},\qquad \d_k=k^2,\qquad k\geq2.\eeq{6.28}

In what follows we will employ the method of induction to
prove Eqs.~(\ref{6.24})--(\ref{6.28}). We start by making
the following ans\"atze for the deformed Dirac operators:
\beq \hat{f}_0^{(k+1)}=\Ds-\radf\left(\a_kN+\b_k\right),\qquad
\hat{f}_0^{\,\prime\,(k+1)}=\Ds-\radf\left(\a'_kN+\b'_k\right),\eeq{6.29}
where $\a_k, \b_k,\a'_k$ and $\b'_k$ are numerical constants. Then, the proof
of Eqs.~(\ref{6.25})--(\ref{6.26}) boils down to finding the following solutions
for these constants:
\beq \a_k=1,\qquad \b_k=-(k+2),\qquad \a'_k=1,\qquad \b'_k=k.\eeq{6.30}
With ans\"atze compatible with Eqs.~(\ref{6.27})--(\ref{6.28}), the PM gauge transformations will also be proved.
Below we present the proofs for $k=0,1,2$, and then for arbitrary $k$.
\newline\vspace{-10pt}\newline
\noindent\underline{$k=0$}: This is the strictly massless case, already considered in Section~\ref{sec:gfgrav}.
Note that because of the $\g$-trace conditions, in writing the involutive system one can replace the deformed divergence
$\Ddiv$ by $\ddiv$. Clearly, Eqs.~(\ref{6.24})--(\ref{6.27}) for $k=0$ take the form of Eqs.~(\ref{5.25})--(\ref{5.27}).
The gauge transformation in this case is given by: $\hat{g}_{-1}=\Dgrad$.\newline\vspace{-10pt}\newline
\noindent\underline{$k=1$}: This corresponds to the simplest nontrivial PM gauge symmetry with depth 2. In this case,
the PM gauge transformation can be implemented by an operator of the form:
\beq \hat{g}_{-2}=\left(\Dgrad\right)^2-\rad \d_1u^2,\qquad \d_1=\text{constant}.\eeq{6.31}
The computation of the gauge variations of the EoM's requires the commutator of $\hat{g}_{-2}$ with
$\{\hat{f}_0^{(2)},\hat{g}_1^{\,\prime},\hat{f}_1\}$, which are given in Eqs.~(\ref{C19})--(\ref{C24}).
These variations simplify to Eqs.~(\ref{C25})--(\ref{C26}) when the involutive system for the gauge parameter
$\ve_{n-2}$ is taken into account. The following choice of constants is required by gauge invariance:
\beq \a_1=1,\qquad \b_1=-3,\qquad \a'_1=1,\qquad \b'_1=1,\qquad \d_1=1.\eeq{6.32}
These coincide with the values given for $k=1$ by Eqs.~(\ref{6.30}) and the gauge transformation~(\ref{6.27}), where
$\d_1$ is precisely the eigenvalue of $(N-n+1)^2$ corresponding to $\ve_{n-2}$.
\newline\vspace{-10pt}\newline
\noindent\underline{$k=2$}: Let us assume that the depth-$3$ gauge transformation is implemented by:
\beq \hat{g}_{-3}=\Dgrad\left[\left(\Dgrad\right)^2-\rad \d_2u^2\right],
\qquad \d_2=\text{constant}.\eeq{6.33}
It is easy to compute the variations of the EoM's of the PM field $\Ps^{(3)}_n$  given the commutation relations~(\ref{C19})--(\ref{C24}). On account
of the involutive system for the gauge parameter $\ve_{n-3}$, these expressions
simplify considerably. Their
explicit forms are given in Eqs.~(\ref{C27})--(\ref{C28}). The vanishing of the gauge variations then requires that
\beq \a_2=1,\qquad \b_2=-4,\qquad \a'_2=1,\qquad \b'_2=2,\qquad \d_2=4,\eeq{6.34}
which are precisely the values Eqs.~(\ref{6.30}) and the gauge transformation~(\ref{6.27}) give for $k=2$. Note that
$\d_2$ is indeed the eigenvalue of $(N-n+1)^2$ corresponding to $\ve_{n-3}$.
The recursion formula~(\ref{6.28}) works too, as it reduces to Eq.~(\ref{6.33}) with $\d_2=4$ for $k=2$.\newline\vspace{-10pt}\newline
\noindent\underline{Generic $k$}: Suppose the involutive system~(\ref{6.24})--(\ref{6.28}) is consistent up to and
including $k=j-2$, for some $j\geq2$. Then, the same system holds good also for $k=j$.
This can be proven with the following ansatz for the depth-$(j+1)$ PM gauge transformation:
\beq \hat{g}_{-j-1}=\left[\left(\Dgrad\right)^2-\rad\d_j u^2\right]\hat{g}_{-j+1},\eeq{6.35}
where $\d_{j}$ is some constant to be determined. Given the ans\"atze~(\ref{6.29}) for the deformed Dirac operators,
it is straightforward to compute the gauge variations of the left-hand sides of the involutive equations for $\Ps_n^{(j+1)}$.
These variations can be written as:
\bea
&\d\!\left[\hat{f}_0^{(j+1)}\Ps_n^{(j+1)}\right]=\left[\Ds-\radf\left(\a_j N+\b_j\right)\right]
\left[\left(\Dgrad\right)^2-\rad\d_j u^2\right]\hat{g}_{-j+1}\ve_{n-j-1},&\nonumber\\
&\d\!\left[\hat{g}_1^{\,\prime}\Ps_n^{(j+1)}\right]=\Ddiv\left[\left(\Dgrad\right)^2-\rad\d_j u^2\right]\hat{g}_{-j+1}\ve_{n-j-1},&\label{6.36}\\
&\d\!\left[\hat{f}_1\Ps_n^{(j+1)}\right]=\ds\left[\left(\dgrad\right)^2-\rad\d_j u^2\right]\hat{g}_{-j+1}\ve_{n-j-1},&\nonumber
\eea{C366666}
where the ``unfree'' gauge parameter $\ve_{n-j-1}$ will be governed by:
\beq \left[\Ds-\radf\left(\a'_jN+\b'_j\right)\right]\ve_{n-j-1}=0,\qquad \Ddiv\,\ve_{n-j-1}=0,\qquad \ds\,\ve_{n-j-1}=0,\eeq{6.37}
given that $\a_{j}$ and $\b_{j}$ and their primed counterparts are some numerical constants.

The right-hand sides of Eqs.~(\ref{6.36}) can be computed by making repeated use of the commutators~(\ref{C19})--(\ref{C24}),
as well as the conditions~(\ref{6.37}) on the gauge parameter. One obtains the following results after a tedious but straightforward
calculation:
\bea
\d\!\left[\hat{f}_0^{(j+1)}\Ps_n^{(j+1)}\right]&=&\left\{\radf\left(\dgrad\right)^{j+1}\mathcal{P}_{j1}+\cdots\right\}\ve_{n-j-1},
\nonumber\\
\d\!\left[\hat{g}_1^{\,\prime}\Ps_n^{(j+1)}\right]&=&\left\{\rad\left(\dgrad\right)^j\mathcal{P}_{j2}+\cdots\right\}\ve_{n-j-1},\label{6.38}\\
\d\!\left[\hat{f}_1\Ps_n^{(j+1)}\right]&=&\left\{\radf\left(\dgrad\right)^j\mathcal{P}_{j3}+\cdots\right\}\ve_{n-j-1},
\nonumber\eea{C388888}
where the ellipses contain lower-derivative terms, and the $\mathcal{P}_j$'s are given by:
\begin{equation}\label{6.39}\begin{split}
\mathcal{P}_{j1}&=(\a'_j-\a_j)N+[\,\b_j'-\b_j-(j+1)(\a_j+1)\,],\\
\mathcal{P}_{j2}&=(j+1)(\a^{\prime\, 2}_j\!-\!1)N^2+(j+1)[\a'_j(2\b'_j+D-1)-2j-D+1]N\\
&~~~+[(j+1)(\b'_j+j+D-1)(\b'_j-j)-2(\d_j-j^2)],\\
\mathcal{P}_{j3}&=(j+1)[(\a'_j-1)N+(\b_j'-j)].
\end{split}\end{equation}

The derivation of the above expressions relies the assumption that the involutive system~(\ref{6.24})--(\ref{6.27}),
and therefore the expressions~(\ref{6.28})--(\ref{6.30}) hold good up to and including $k=j-2$. Now, vanishing of the gauge
variations~(\ref{6.38}) necessarily requires that the gauge parameter $\ve_{n-j-1}$ belongs simultaneously to the kernels of $\mathcal{P}_{j1}$, $\mathcal{P}_{j2}$ and $\mathcal{P}_{j3}$. For a non-trivial gauge parameter, however, such conditions
can be satisfied iff the operators themselves vanish. This leads to a unique
set of solutions for $\d_j, \a_j, \b_j, \a'_j$ and $\b'_j$, which coincides with that spelled out by Eqs.~(\ref{6.28})
and~(\ref{6.30}) for $k=j$. That these values are also sufficient for the gauge variations~(\ref{6.38}) to vanish can be
proved by explicitly showing that all the lower-derivative terms vanish. It is not difficult to convince oneself of this
fact for any given $j$, but we conclude without presenting this tedious exercise.

We now summarize our results. In AdS space, the involutive system of a rank-$n$ depth-$(k+1)$ PM fermion $\Pspm$ and its
rank-$(n-k-1)$ gauge parameter $\vepm$ reads:
\begin{equation}\label{6.40}\begin{split}
\left(\nbs-\mathfrak{m}_k\right)\Pspm&=0,\qquad \ddiv\,\Pspm=0,\qquad \ds\,\Pspm=0,\\
\left(\nbs-\mathfrak{m}'_k\right)\vepm&=0,\qquad \ddiv\,\vepm=0,\qquad \ds\,\vepm=0,
\end{split}\end{equation}
with the mass terms given for $k\geq0$ by Eqs.~(\ref{6.26}). Note that the above system has been presented without the
$\ds$-piece appearing in the deformed divergence $\Ddiv$; this possible because the $\g$-trace conditions themselves
are included in the system~(\ref{6.40}). The depth-$(k+1)$ PM gauge transformations of this involutive system are of the form:
\beq \d\Pspm=\hat{g}_{-k-1}\vepm,\eeq{6.41}
where $\hat{g}_{-k-1}$ is spelled out in Eq.~(\ref{6.27}), and it contains up to $(k+1)$ derivatives.

\section{Lie Algebra of Operators}\label{sec:Consistency}

This section studies the Lie superalgebra formed by the various operators acting on symmetric
tensor(-spinor)s in maximally symmetric spaces. Section~\ref{sec:flatalg} presents the flat-space
algebra, while Section~\ref{sec:adsalg} shows how in AdS space the algebra closes only nonlinearly with
a central extension. In this regard, let us note that nonlinear Lie algebras\footnote{They appear in
Physics as Higgs algebra~\cite{Higgs:1978yy} and $W_3$ algebra~\cite{Schoutens:1989tn}, in
quantum optics~\cite{Beckers:1998ef}, and so on.} are generalizations of ordinary Lie algebras containing
different order products of the generators on the right-hand side of the defining brackets without violating Jacobi identities. In AdS space, the nonlinear bosonic subalgebra of operators has been studied in~\cite{Hallowell:2005np,Buch3,Hallowell:2007qk,Nutma:2014pua,Rahman:2016tqc}, while the full supersymmetric algebra was
considered in~\cite{Hallowell:2005np,Hallowell:2007qk,Buch4}.

\subsection{Algebra in Flat Space}\label{sec:flatalg}

In flat space, the Lie superalgebra of all the operators on symmetric tensor(-spinor)s turns out to
be a subalgebra of $osp(4|1)$, whereas the Lie subalgebra formed only by the bosonic generators is a subalgebra
of $sp(4)$~\cite{Hallowell:2007qk}. In order to present the Lie algebras, let us first list all the flat-space
operators, along with their various properties (Table~3).
\begin{table}[ht]
\caption{Operators on Symmetric Tensor(-Spinor)s in Flat Space}
\vspace{5pt}
\centering
\begin{tabular}{c c c c c}
\hline\hline
Operator~~~&~~~Symbol~~~&~~~Definition~~&~~Weight $(w)$~~&Type\\ [0.5ex]
\hline
d'Alembertian & $\dal$ & $\de^2$ & $~~0$ & \\
Divergence & $g_1$ & $\div$ & $-1$ & \\
Symmetrized Gradient& $g_{-1}$ & $\grad$ & $+1$ & bosonic\\
Trace & $g_2$ & $d^2$ & $-2$ & \\
Symmetrized Metric& $g_{-2}$ & $u^2$ & $+2$ & \\
\hline
Massless Dirac & $\dir$ & $\des$ & $~~0$ & \\
Gamma Trace & $f_1$ & $\ds$ & $-1$ & fermionic\\
Symmetrized Gamma & $f_{-1}$ & $\us$ & $+1$ & \\
\hline
Index Operator& $N$ & $\spin$ & $~~0$ & bosonic\\
\hline\hline
\end{tabular}
\end{table}
\vspace{5pt}

\noindent
Note that in the above list we have included, among other things, all the operators that appear in the EoM's
of symmetric tensors and tensor-spinors, namely $\{\dal,g_1,g_2,\dir,f_1\}$. However, it also includes the
hermitian conjugates (in the sense of footnote~5) of these operators as well: $\{\dal,g_{-1},g_{-2},\dir,f_{-1}\}$.
The positive-weight operators appear not in the EoM's, but in the hermitian conjugates thereof; their inclusion
is tantamount to admitting a Lagrangian formulation, e.g., via BRST approach~\cite{Buch3,Buch4}. Last but not
the least, the index operator $N$ is added as it provides a grading to all the operators.

The graded commutators of all these operators are given in Table~4. The computation is easy because
ordinary derivatives commute: $[\de_\m,\de_\n]\F=0=[\de_\m,\de_\n]\Ps$. In particular,
$[\de_\m,\de_\n]$ is blind to the statistical nature of the field. As we will see in the next section,
this seemingly naive observation provides valuable input when it comes to curved backgrounds.
\vspace{-15pt}
\begin{table}[ht]
\caption{Graded Commutators of Flat-Space Operators}
\vspace{5pt}
\centering
\begin{tabularx}{0.95\textwidth} {
  | >{\centering\arraybackslash}X
  | >{\centering\arraybackslash}X
  | >{\centering\arraybackslash}X
  | >{\centering\arraybackslash}X
  | >{\centering\arraybackslash}X
  | >{\centering\arraybackslash}X
  | >{\centering\arraybackslash}X
  | >{\centering\arraybackslash}X
  | >{\centering\arraybackslash}X
  | >{\centering\arraybackslash}X | }
 \hline\hline
 $[\downarrow,\!\rightarrow\!\}$ & $N$& $\dal$ & $g_1$ & $g_{-1}$& $g_2$& $g_{-2}$& $\dir$& $f_1$& $f_{-1}$\\
 \hline
 $N$  & $0$ & $0$ & $-1$ & $+1$ & $-2$ & $+2$ & $0$ & $-1$ & $+1$ \\
 \hline
 $\dal$  & \phantom  & $0$ & $0$ & $0$ & $0$ & $0$ & $0$ & $0$ & $0$ \\
 \hline
 $g_1$  & \phantom   & \phantom  & $0$ & $\dal$ & $0$ & $2g_{-1}$ & $0$ & $0$ & $\dir$ \\
 \hline
 $g_{-1}$  & \phantom   & \phantom  & \phantom  & $0$ & $-2g_1$ & $0$ & $0$ & $-\dir$ & $0$ \\
 \hline
 $g_2$  & \phantom   & \phantom  & \phantom  & \phantom  & $0$ & $\scriptstyle{4N+2D}$ & $0$ & $0$ & $2f_1$ \\
 \hline
 $g_{-2}$  & \phantom   & \phantom  & \phantom  & \phantom  & \phantom  & $0$ & $0$ & $-2f_1$ & $0$ \\
 \hline
 $\dir$  & \phantom  & \phantom  & \phantom  & \phantom  & \phantom  & \phantom  & $2\dal$ & $2g_1$ & $2g_{-1}$ \\
 \hline
 $f_1$  & \phantom  & \phantom  & \phantom  & \phantom  & \phantom  & \phantom  & \phantom  & $2g_2$ & $\scriptstyle{2N+D}$ \\
 \hline
 $f_{-1}$  & \phantom   & \phantom  & \phantom  & \phantom  & \phantom  & \phantom  & \phantom  & \phantom  & $2g_{-2}$ \\
 \hline\hline
\end{tabularx}
\end{table}

\subsection{Algebra in AdS Space}\label{sec:adsalg}

In a curved background, the deformed counterparts of the flat-space operators
in Table~3 do not form an algebra in general because of non-commutativity of
covariant derivatives. It can be shown that the bosonic subalgebra may close, perhaps nonlinearly, only in constant curvature manifolds~\cite{Buchbinder:2011vp},
or in Freund-Rubin type backgrounds $\text{AdS}_p\times\text{S}^q$ with equal radii~\cite{Rahman:2016tqc},
in which case the algebra is simply a covariant uplift of the $\text{AdS}_p$ algebra.

In the supersymmetric case, however, there is an immediate puzzle in deforming the flat-space generators: the
commutator of covariant derivatives acts differently on bosonic and fermionic fields, as wee see from Eq.~(\ref{C3}).
Then, how can the same operator algebra be realized on states with different statistics? The resolution of the puzzle
lies in that a central charge $Z$ must be introduced in the following way. In AdS space, when Eq.~(\ref{C3}) is compared
with Eqs.~(\ref{6.23}), the following possibility immediately comes to one's mind:
\beq \D_\m\equiv\nb_\m+Z\g_\m,\qquad \text{such that}\quad[\D_\m,\D_\n]=\left\{
                                                    \begin{array}{ll}
                                                      -\tfrac{2}{L^2}u_{[\m}d_{\n]}, & \hbox{for bosons;} \\
                                                      -\tfrac{2}{L^2}u_{[\m}d_{\n]}, & \hbox{for fermions,}
                                                    \end{array}
                                                  \right.
\eeq{7.2}
where $Z$ is a bosonic operator of mass dimension 1 that commutes with all the other generators.
A bosonic state $\F$ and a fermionic state $\Ps$ carry different charges under $Z$:
\beq Z\F=0,\qquad Z\Ps=-\tfrac{1}{2L}\Ps.\eeq{7.3}
In other words, deformed covariant derivative $\D_\m$ in the supersymmetric case
reduces to $\nb_\m$ and $\left(\nb_\m-\tfrac{1}{2L}\g_\m\right)$ respectively for bosons and fermions. As
a supersymmetric generalization of~(\ref{C20.00}), one has the commutation relation:
$[\g^\m,\D^\n]=[\D^\m,\g^\n]=2Z\g^{\m\n}$.

In what follows we will set the AdS radius to unity: $L=1$. One can start by defining the following
deformed bosonic operators:
\begin{equation}\label{7.4} \begin{split}
\text{Divergence}:~\mg_1&\equiv\Ddiv,\\
\text{Symmetrized Gradient}:~\mg_{-1}&\equiv\Dgrad,\\
\text{d'Alembertian}:~\mg_0&\equiv[\mg_1,\mg_{-1}].
\end{split}\end{equation}
In view of Eqs.~(7.1), the deformed d'Alembertian $\mg_0$ can also be expressed as:
\beq \mg_0=\D^2-N(N+D-2)+u^2d^2.\eeq{7.5}
Furthermore, the deformed Dirac operator can be chosen such that
its anti-commutation relations with the other fermionic operators mimic
their flat-space counterparts. It is easy to check that the following choices
achieve the desired feat:
\begin{equation}\label{7.6} \begin{split}
\text{Dirac}:~\mf_0&\equiv\Ds-(D-1)Z,\\
\text{Gamma Trace}:~\mf_1&\equiv\g\!\cdot\!d,\\
\text{Symmetrized Gamma}:~\mf_{-1}&\equiv\g\!\cdot\!u.
\end{split}\end{equation}
The remaining three bosonic operators include the index operator
$N\equiv u\!\cdot\!d$, and
\begin{equation}\label{7.7} \begin{split}
\text{Trace}:~\mg_2&\equiv d^2,\\
\text{Symmetrized Metric}:~\mg_{-2}&\equiv u^2.
\end{split}\end{equation}
This exhausts the list of operators. It is straightforward to calculate all the graded commutators. While many of them
close linearly like their flat-space counterparts, nonlinearity arises in some of the commutators. The results are summarized
below in Table~5.
\begin{table}[ht]
\caption{Graded Commutators of Operators in AdS}
\vspace{5pt}
\centering
\begin{tabularx}{0.95\textwidth} {
  | >{\centering\arraybackslash}X
  | >{\centering\arraybackslash}X
  | >{\centering\arraybackslash}X
  | >{\centering\arraybackslash}X
  | >{\centering\arraybackslash}X
  | >{\centering\arraybackslash}X
  | >{\centering\arraybackslash}X
  | >{\centering\arraybackslash}X
  | >{\centering\arraybackslash}X
  | >{\centering\arraybackslash}X
  | >{\centering\arraybackslash}X | }
 \hline\hline
 $\![\downarrow,\!\rightarrow\!\}\!$ & $N$& $\mg_0$ & $\mg_1$ & $\mg_{-1}$& $\mg_2$& $\mg_{-2}$& $\mf_0$& $\mf_1$& $\mf_{-1}$& $Z$\\
 \hline
 $N$  & $0$ & $0$ & $-1$ & $+1$ & $-2$ & $+2$ & $0$ & $-1$ & $+1$& $0$ \\
 \hline
 $\mg_0$  & \phantom  & $0$ & $\mc_{23}$ & $\mc_{24}$ & $0$ & $0$ & $\mc_{27}$ & $\mc_{28}$ & $\mc_{29}$ & $0$ \\
 \hline
 $\mg_1$  & \phantom   & \phantom  & $0$ & $\mg_0$ & $0$ & $2\mg_{-1}$ & $\mc_{37}$ & $0$ & $\mc_{39}$ & $0$ \\
 \hline
 $\mg_{-1}$  & \phantom   & \phantom  & \phantom  & $0$ & $-2\mg_1$ & $0$ & $\mc_{47}$ & $-\mc_{39}$ & $0$ & $0$ \\
 \hline
 $\mg_2$  & \phantom   & \phantom  & \phantom  & \phantom  & $0$ & $\scriptstyle{4N+2D}$ & $0$ & $0$ & $2\mf_1$ & $0$ \\
 \hline
 $\mg_{-2}$  & \phantom   & \phantom  & \phantom  & \phantom  & \phantom  & $0$ & $0$ & $-2\mf_1$ & $0$ & $0$ \\
 \hline
 $\mf_0$  & \phantom  & \phantom  & \phantom  & \phantom  & \phantom  & \phantom  & $\mc_{77}$ & $2\mg_1$ & $2\mg_{-1}$ & $0$ \\
 \hline
 $\mf_1$  & \phantom  & \phantom  & \phantom  & \phantom  & \phantom  & \phantom  & \phantom  & $2\mg_2$ & $\scriptstyle{2N+D}$ & $0$ \\
 \hline
 $\mf_{-1}$  & \phantom   & \phantom  & \phantom  & \phantom  & \phantom  & \phantom  & \phantom  & \phantom  & $2\mg_{-2}$ & $0$ \\
\hline
 $Z$  & \phantom   & \phantom  & \phantom  & \phantom  & \phantom  & \phantom  & \phantom  & \phantom  & \phantom & $0$ \\
 \hline\hline
\end{tabularx}
\end{table}

In particular, the deformed d'Alembertian $\mg_0$ has nonlinear commutation relations with the divergence
and gradient as well as with all the fermionic operators:
\bea
&[\mg_0,\mg_1]=2(2N+D-1)\mg_1-4\mg_{-1}\mg_2\equiv\mc_{23},&\nonumber\\
&[\mg_0,\mg_{-1}]=-2\mg_{-1}(2N+D-1)+4\mg_{-2}\mg_1\equiv\mc_{24},\nonumber\\
&[\mg_0,\mf_0]=2\left(\mf_{-1}\mg_1-\mg_{-1}\mf_1\right)\equiv\mc_{27},&\label{7.8}\\
&[\mg_0,\mf_1]=(2N+D-1)\mf_1-2\mf_{-1}\mg_2+4Z\left(\mg_1-\mf_1\mf_0\right)
\equiv\mc_{28},\nonumber&\\
&[\mg_0,\mf_{-1}]=-\mf_{-1}(2N+D-1)+2\mg_{-2}\mf_1
-4Z\left(\mg_{-1}-\mf_0\mf_{-1}\right)\equiv\mc_{29}.&\nonumber\eea{7.88888}
The Dirac operator $\mf_0$  also closes nonlinearly with the divergence, gradient, and itself:
\bea &[\mf_0,\mg_1]=(N-D+1)\mf_1-2(D-1)Z^2\mf_1-\mf_{-1}\mg_2+2Z\left(\mg_1-\mf_1\mf_0\right)\equiv\mc_{37},&
\nonumber\\
&[\mf_0,\mg_{-1}]=-\mf_{-1}(N-D+1)+2(D-1)Z^2\mf_{-1}+\mg_{-2}\mf_1-2Z\left(\mg_{-1}-\mf_0\mf_{-1}\right)\equiv\mc_{47},&
\label{7.9}\\
&\{\mf_0,\mf_0\}=2\mg_0+2N(N+D-1)-2\left(\mg_{-2}\mg_2+\mf_{-1}\mf_1\right)+2(D-1)^2Z^2\equiv\mc_{77}.&
\nonumber\eea{7.99999}
Last but not the least, we have nonlinear closure of the following commutators:
\beq [\mg_1,\mf_{-1}]=-[\mg_{-1},\mf_1]=\mf_0+Z\left(2N+D-1-2\mf_{-1}\mf_1\right)\equiv\mc_{39}.\eeq{7.10}

Some comments are in order at this point. First, the AdS nonlinear superalgebra (Table~5) contains a bosonic
central charge $Z$, which does not show up in the flat-space Lie superalgebra (Table~4). The appearance of a
central charge in AdS, when fermionic fields are considered, was already noted in~\cite{Buch4}. This central
extension is however not required when one considers only symmetric tensors in AdS~\cite{Rahman:2016tqc}, i.e.,
for the bosonic algebra generated by $\{\mg_0,\mg_1,\mg_{-1},\mg_2,\mg_{-2},N\}$. Second, one can perform a
covariant uplift of the $\text{AdS}_D$-superalgebra to render it consistent for any Freund-Rubin type background
$\text{AdS}_p\times\text{S}^{q}$ with equal radii, exactly the same way the bosonic algebra
can be~\cite{Rahman:2016tqc}. In this case, the $\text{AdS}_p\times\text{S}^{q}$-superalgebra will be non-analytic
in the neighborhood of flat space.

\section{Conclusions}\label{sec:remarks}

In this article, we have studied the involutive systems of equations describing the free propagation of
massive, massless and partially massless
symmetric tensors and tensor-spinors. For massive and massless fields, we have employed the involutive deformation
method to find consistent dynamical equations and constraints/gauge-fixing conditions, compatible with gauge symmetries
if present, in gravitational and electromagnetic backgrounds. For partially massless fields, we have given explicit
expressions for the gauge transformations and mass parameters at arbitrary depth. We have also shown that the Lie
superalgebra of operators acting on symmetric tensor(-spinor)s in AdS space closes nonlinearly as an extension of
the flat-space algebra by a bosonic central charge.

As pointed out in the Introduction, in the involutive approach, all the consistency issues regarding the
propagation of higher-spin fields are under proper control. The mutual compatibility and possible gauge invariance of the
equations describing the system are taken care of by the involutive structure itself, which thereby preserves the
degrees of freedom count. On the other hand, higher-derivative terms may inflict Ostrogradsky
instability~\cite{Ostrogradksi}, while non-canonical kinetic terms may affect hyperbolicity or causal propagation.
The latter issues become manifest in the involutive approach, unlike in the Lagrangian formulation, so much so that
avoiding them simply becomes a matter of choice. More importantly, the involutive deformation method can also
be employed to construct consistent interactions~\cite{KaLySh3}. This goes beyond the scope of our present work.

The various deformed involutive systems presented throughout this article could be viewed as the infrared
limits of some effective-field-theory equations. Let us recall from Sections~\ref{sec:mb} and~\ref{sec:mf} that,
for higher-curvature and higher-derivative terms in the equations, the suppression scales $\L$ and $\bar{\L}$ were
introduced. For a given system, such a scale ought to be parametrically larger than other mass scales
in order for an effective field theory description to be valid. Eventually, for the sake of simplicity, we considered
only the infrared limit by sending these scales to infinity. This also rids the systems of higher derivatives and/or
kinetic deformations that might otherwise jeopardize causal propagation. One could however keep these scales finite,
and move on to searching for the deformed involutive systems. Thus, one would find higher-curvature corrections
to the equations of motion, e.g., those for massive higher-spin fields in string
theory~\cite{AN,Klishevich,PRS,Buchbinder:1999be}.

We only considered the propagation of a \emph{single} higher-spin field in a \emph{pure}
gravitational or electromagnetic background. One could generalize the analysis for interactions with more generic
backgrounds~\cite{Cortese:2017ieu}, and thus find yes-go results. For example, as already mentioned in
Section~\ref{sec:gfem}, Einstein-Maxwell backgrounds do admit the propagation of a charged spin-$\tfrac{3}{2}$
gauge field. The assumption of a field in isolation is a strong one since, in a nontrivial
background, various fluctuations of different spins may mix in the EoM's even at the linear level. Relaxing this
assumption would again lead to yes-go results by weakening the constraints on the backgrounds, otherwise required
by consistency. One obvious example includes the graviton fluctuation in any geometry sourced by a nontrivial
stress-energy tensor. Surely, its propagation will be consistent, thanks to General Relativity, but the linearized
equations will inevitably mix the graviton with the fluctuations of the fields contributing to the stress-energy tensor.
On the other hand, when gravity is dynamical, any finite number of massive higher-spin fields could lead to
causality violation~\cite{Kundu1,Kundu2}.

By construction, the involutive deformations we obtained have smooth flat limits. Accordingly, so do the deformed
masses chosen in Sections~\ref{sec:mb} and~\ref{sec:mf}; the deformations however are non-unique in
that they could be arbitrary polynomials of the index operator $N$. For gravitational backgrounds, these ambiguities
could be removed by requiring smooth massless limits. However, the non-uniqueness of mass deformations persists in the
case of electromagnetic backgrounds. In fact, it is even consistent to start with flat-space masses that are polynomials
of $N$, generalizing the Regge law in string theory.

What r\^ole would mixed-symmetry fields play if included in the spectrum? Let us recall that even a massive higher-spin
fermion calls for an AdS background. While $\text{AdS}_{10}$ is not a solution of
superstring theory, $\text{AdS}_5\times\text{S}^5$ is. As noted in Section~\ref{sec:adsalg}, one can perform
a covariant uplift of the higher-spin involutive systems to make them consistent even in such a
background~\cite{Rahman:2016tqc}. In the latter case, however, the deformations will not be analytic in the neighborhood
of flat space~\cite{Rahman:2016tqc}. This is in sharp contrast with string theory. While our analysis is restricted to
symmetric tensor(-spinors)s only, it is the mixed-symmetry fields in string theory that ensure analyticity in the background
curvature. This point could be further justified by considering the theory of charged open bosonic strings in a background gauge
field~\cite{AN,PRS}. The full Virasoro algebra ensures consistent propagation of the string fields. However, if the subleading
Regge trajectories are excluded by switching off some of the oscillators, the remaining non-trivial generators no longer form
an algebra~\cite{Rahman:2016tqc}.

\newpage
\subsection*{Acknowledgments}

We would like to thank S.~Biswas for initial collaboration, and I.~Cortese, K.~Mkrtchyan, M.~Sivakumar, Z.~Skvortsov, and M.~Taronna for
valuable comments. RR acknowledges the kind hospitality and support of the Erwin Schr\"odinger International Institute for Mathematics and
Physics and the organizers of the scientific activity ``Higher Spins and Holography'' (March 11--April 05, 2019),
during which part of this work was presented.

\begin{appendix}
\numberwithin{equation}{section}

\section{Involutive System of Equations}\label{sec:appendix}

Involutive systems of partial differential equations (PDE) and how they control the number
of DoF's of a dynamical system are well studied in the literature~\cite{Inv-book}. Related to the
count of Cauchy data~\cite{Inv-book}, the DoF count can be made by relying on the notion of ``strength''
of an involutive system. This direction was first explored by Einstein~\cite{Boss}, and further
developed by subsequent authors~\cite{inv1,inv2,inv3,inv4}. In this appendix, we explain the basics
of involution and derive some necessary formulae for DoF count. For technical details,
which we will skip, readers may resort to Ref.~\cite{KaLySh3} and references therein.

Let us work with the convention that repeated indices appearing all as either covariant or contravariant
ones are symmetrized with minimum number of terms. This gives us the rules:
$\m(k)\m=\m\m(k)=(k+1)\m(k+1)$,~$\m(k)\m(2)=\m(2)\m(k)=\binom{k+2}{2}\,\m(k+2)$,
$\m(k)\m(k')=\m(k')\m(k)=\binom{k+k'}{k}\,\m(k+k')$, and so on, where $\m(k)$ has a unit weight by convention,
and so the proportionality coefficient gives the weight of the right hand side.

\subsection{Involution Basics}\label{sec:invbas}

We consider a set of fields $\Phi^A$, with $A=1,2,...,f$, and denote their $k$-th space-time derivative
by $\Phi^A_{\m(k)}$. Let their dynamics be described by the following system of PDE's:
\beq T^a[\Phi^A,\Phi_\m^A,\ldots,\Phi_{\m(p)}^A]=0,\quad\text{with}\quad a=1,2,...,t.\eeq{eq:eom}
The maximal derivative order $p$ is called the \emph{order} of the system. Consider any order-$p'$
subsystem: $T^b[\Phi^A,\de_\m\Phi^A,\ldots,\Phi_{\m(p')}^A]=0,\,b\subset a,\,p'\leq p$.
The system~(\ref{eq:eom}) is \emph{involutive} if it contains all the differential consequences of
$\text{order}\leq p'$ derivable from the subsystem.

If the system~(\ref{eq:eom}) is involutive, it may possess nontrivial identities of the form:
\beq \sum_a L_a^i T^a=0,\qquad i=1,2,\dots,l,\eeq{eq:lgen}
with $L^i_a$ being local differential operators. These are called the \emph{gauge identities}. The
(total) \emph{order} of a gauge identity is again the maximal derivative order appearing therein.
Note that gauge identities are more generic than Noether identities, and may exist even without gauge
symmetries. The two coincide only for a set of Lagrangian equations that is involutive to begin
with~\cite{KaLySh3}. Gauge identities play an important r\^ole in that they reflect algebraic
consistency of the involutive system, and control the DoF count.

In general, the involutive system~(\ref{eq:eom}) may also enjoy local \emph{gauge symmetries}:
\beq \d_\ve\Phi^A=\sum_\a R_\a^A\ve^\a,\qquad\d_\ve T^a|_{\,T=0}=0,\qquad \a=1,2,\dots,r,\eeq{gaugesymm}
where $\ve^\a$ are the gauge parameters, while $R_\a^A$ are differential operators of finite order.
It may happen that the gauge parameters are not arbitrary (as is often the case with partial gauge fixing),
but they themselves are governed by an involutive system of equations. In the bulk of the article, we only
have to deal with gauge symmetries of the latter kind.

\subsection{DoF Count}\label{sec:invdof}

Let us assume that $\Phi^A(x)$ are analytic functions of the space-time coordinates $x^\m$.
One may write down a Taylor series expansion of $\Phi^A(x)$ around some point $x_0^\m$:
\beq \Phi^A(x)=\bar\Phi^A+\sum_{k=1}^\infty\frac{1}{k!}\,\bar\Phi_{\m_1\cdots\m_k}^A(x-x_0)^{\m_1}
\cdots(x-x_0)^{\m_k},\eeq{A1}
where a ``bar'' stands for the corresponding unbarred quantity evaluated at $x=x_0$.
Here, the Taylor coefficients at $\mathcal{O}(k)$ are furnished by the quantities $\bar\Phi_{\m(k)}^A$,
which constitute a set of monomials. Because of the EoM's~(\ref{eq:eom}), however, not all of these
monomials remain undetermined. Moreover, if the system enjoys gauge symmetries, some
of the monomials will be physically equivalent. Let us define the following quantities:
\bea n_k&=&\text{Total number of monomials at}~\mathcal{O}(k),\nonumber\\
\hat{n}_k&=&\text{Number of undetermined gauge-inequivalent monomials at}~\mathcal{O}(k).\nonumber
\eea{}
Then, the number of physical DoF per point in $D$ dimensions will be given by:
\beq\mathfrak{D}~=~\frac{f}{2(D-1)}\lim_{k\rightarrow\infty}\left(k\,\frac{\hat{n}_k}{n_k}\right).\eeq{A2}
This formula measures the number of physical DoF's as the proliferation of the physical
monomials relative to the unconstrained ones, \`a la Einstein~\cite{Boss}. For large $k$, we will see below
that $\hat{n}_k\sim n_k/k$, and so the above limit yields a finite number. The dimension-dependent
proportionality factor can be obtained, for example, by matching with the DoF count for a scalar field.
Note that the formula~(\ref{A2}) gives the number of physical polarizations, i.e., the number of physical
DoF's in configuration space.

We will make use of Eq.~(\ref{A2}) for a system of free-field equations. In other words, the EoM's~(\ref{eq:eom})
are assumed to be linear in the fields. At $x=x_0$, they can be written as:
\beq \mathcal{J}^{a,\,\n(p)}_A\bar\Phi_{\n(p)}^A+\mathfrak{b}^a=0,\qquad
\mathcal{J}^{a,\,\n(p)}_A\equiv\frac{\d T^a}{\d \Phi_{\n(p)}^A},\eeq{A3}
where $\mathfrak{b}^a$ will be linear in $\bar\Phi_{\n(p')}^A$ with $p'<p$. Note that the quantity
$\mathcal{J}^{a,\,\n(p)}_A$ is called the \emph{zeroth-order symbol matrix}. In general, one may have the
\emph{m$^{\,th}$-order symbol matrix}:
\beq \mathcal{J}^{a,\,\n(k)}_{A,\,\m(m)}\equiv\frac{\d T^a_{\m(m)}}{\d \Phi_{\n(k)}^A},\qquad
m\equiv k-p\geq0,\eeq{A4}
where $T^a_{\m(m)}$ denotes the $m$-th gradient of the EoM's. Then, the $m$-th gradient of
Eq.~(\ref{eq:eom}) evaluated at $x=x_0$ gives a straightforward generalization of~(\ref{A3}),
which is
\beq \mathcal{J}^{a,\,\n(k)}_{A,\,\m(m)}\bar\Phi_{\n(k)}^A+~\cdots~=~0,\eeq{A5}
where the ellipses stand for linear terms in the monomials $\bar\Phi_{\n(k')}^A$ at order $k'<k=p+m$.
The above equation involves monomials at order $k\geq p$; their total number is given by:
\beq n_k=f{k+D-1\choose k}.\eeq{A5.1}
The space of these monomials is determined by the finite system~(\ref{A5}) of linear inhomogeneous
equations, whose total number amounts to
\beq n^T_k=t{m+D-1\choose m}=t{k-p+D-1\choose k-p}.\eeq{A5.2}

Note that in order for the system~(\ref{A5}) to be compatible, a left null vector of
the symbol matrix must annihilate the inhomogeneous term, and vice versa.
This compatibility criterion is automatically satisfied by any involutive system (since otherwise the
system would not be involutive in the first place). Existence of a left null vector of the $m^{th}$-order
symbol matrix then gives rise to an identity at $\mathcal{O}(k)$. Such an identity must be a consequence
of the gauge identities~(\ref{eq:lgen}). If $q$ is the total order of the gauge identities, then taking
$(k-q)$-th gradient of Eq.~(\ref{eq:lgen}) leads us to an identity of the following form:
\beq \Theta_{a,\,\m(k-q)}^{i,\,\n(m)}\mathcal{J}^{a,\,\r(k)}_{A,\,\n(m)}\bar\Phi_{\r(k)}^A
+~\cdots~=~0,\qquad k\geq q\geq p,\qquad m=k-p\geq0,\eeq{A6}
where the ellipses contain terms linear in $\bar\Phi_{\n(k')}^A$ with $k'<k$. Because $k$ can be made
arbitrarily large, in order for identity~(\ref{A6}) to hold good, it is necessary that
\beq \Theta_{a,\,\m(k-q)}^{i,\,\n(m)}\mathcal{J}^{a,\,\r(k)}_{A,\,\n(m)}=0,\qquad\text{for large}~k.\eeq{A7}
Therefore, $\Theta_{a,\,\m(k-q)}^{i,\,\n(m)}$ serves as a set of left null vectors of the symbol matrix
$\mathcal{J}^{a,\,\r(k)}_{A,\,\n(m)}$ for large $k$. The total number of these null vectors is equal to
\beq n^L_k=l{k-q+D-1\choose k-q}.\eeq{A7.1}
They will be linearly independent if the original gauge identities~(\ref{eq:lgen}) are \underline{irreducible}.

The number of $\mathcal{O}(k)$-monomials determined by the system is given by the rank of the symbol
matrix of order $m\!=\!k-p$. The rank, in turn, is the difference between the number~(\ref{A5.2})
of $\mathcal{O}(k)$-equations and the number of \underline{independent} left null vectors of the symbol
matrix. Once these quantities are known, one can count the number of undetermined $\mathcal{O}(k)$-monomials.
The DoF count further requires modding out gauge-equivalent monomials if gauge symmetries are present
in the system.

Let us Taylor expand the local gauge symmetry parameters appearing in Eq.~(\ref{gaugesymm}):
\beq \ve^\a(x)=\bar\ve^{\,\a}+\sum_{k=1}^\infty\frac{1}{k!}\,\bar\ve^{\,\a}_{\m_1\cdots\m_k}(x-x_0)^{\m_1}
\cdots(x-x_0)^{\m_k}.\eeq{A8}
If $s$ is the \emph{order} of the gauge transformation (maximal order of $R_\a^A$), then taking
$m$-th gradient of the equation: $\d_\ve T^a|_{\,T=0}=0$, leads us to the following schematic form:
\beq \mathcal{J}^{a,\,\n(k)}_{A,\,\m(m)}\left\{\Xi_{\a,\,\n(k)}^{A,\,\r(k+s)}\bar\ve^{\,\a}_{\r(k+s)}\right\}
+~\cdots~=~0,\qquad m=k-p\geq0,\eeq{A9}
where the ellipses contain terms linear in $\bar\ve_{\n(k')}^{\,\a}$ with $k'<k+s$. Again, since $k$ can be
arbitrarily large, Eq.~(\ref{A9}) necessarily implies the following\footnote{If the gauge parameters are
completely arbitrary, which is not the case we deal with in this article, the relation would be true for
any $k=p+m$.}:
\beq \mathcal{J}^{a,\,\n(k)}_{A,\,\m(m)}\,\Xi_{\a,\,\n(k)}^{A,\,\r(k+s)}=0,\qquad\text{for large}~k=p+m.\eeq{A10}
Therefore, $\Xi_{\a,\,\n(k)}^{A,\,\r(k+s)}$ furnishes a set of right null vectors of the $m^{th}$-order
symbol matrix for large $k$. The total number of such right null vectors is given by:
\beq n^R_k=r{k+s+D-1\choose k+s}.\eeq{A11}
These vectors will all be nontrivial and linearly independent for \underline{irreducible} gauge symmetries
with \underline{unconstrained} parameters. If it is otherwise, the DoF count becomes more involved. This is
also the case when the gauge identities are reducible. Taking such cases into account, we will now derive
some formulae for DoF count.

\subsubsection{No Gauge Symmetries}\label{sec:nogauge}

In general, the system~(\ref{eq:eom}) may contain equations of various orders. Suppose the number of equations
at order $p$ is given by $t_p$. The generalization of the count~(\ref{A5.2}) would read:
\beq n^T_k=\sum_{p}t_p{k-p+D-1\choose k-p}.\eeq{A12}
The gauge identities may come at different orders as well. Moreover, the gauge identities may not be irreducible.
Suppose there are $l_{q,j}$ number of gauge identities at total order $q$ and reducibility order $j$. It is not
difficult to convince oneself that the generalization of~(\ref{A7.1}) to the total count of \underline{independent}
gauge identities will be given by:
\beq n^L_k=\sum_{q,j}(-)^jl_{q,\,j}{k-q+D-1\choose k-q}.\eeq{A13}
In the absence of gauge symmetries, the number of undetermined physical monomials at $\mathcal{O}(k)$ will be
given by: $\hat{n}_k=n_k-(n_k^T-n_k^L)$, which is equal to
\beq \hat{n}_k~=~f{k+D-1\choose k}-\sum_n\left(t_n-\sum_j(-)^jl_{n,\,j}\right){k-n+D-1\choose k-n}.
\eeq{A14}
We can make use of the following asymptotic expansion for binomial coefficients~\cite{inv1,inv4}:
\beq {k\pm n+D-1\choose k\pm n}={k+D-1\choose k}
\left\{1\pm\frac{n}{k}(D-1)+\mathcal{O}\left(\frac{1}{k^2}\right)\right\},\qquad k\rightarrow\infty.\eeq{A15}
Now, plugging the above expansion into Eq.~(\ref{A14}) and dividing by Eq.~(\ref{A5.1}), we obtain:
\beq\frac{f\hat{n}_k}{n_k}~=~\mc+\left(\frac{D-1}{k}\right)\sum_nn\left(t_n-\sum_j(-)^jl_{n,\,j}\right)
+\mathcal{O}\left(\frac{1}{k^2}\right),\eeq{A16}
where $\mc$ is called the \emph{compatibility coefficient}, given by:
\beq \mc\equiv f-\sum_n\left(t_n-\sum_j(-)^jl_{n,\,j}\right).\eeq{A17}
We will assume that the system~(\ref{eq:eom}) is \emph{absolutely compatible}, i.e., $\mc=0$.
In this case, the DoF count~(\ref{A2}) can be computed by taking a limit of Eq.~(\ref{A16}),
which gives:
\beq \mathfrak D=\tfrac{1}{2}\sum_{n}n\left(t_n-\sum_j(-)^jl_{n,\,j}\right).\eeq{A18}
This is the formula for physical DoF count of an absolutely compatible involutive system of with
reducible gauge identities, but no gauge symmetries.

\subsubsection{Irreducible Gauge Symmetries with Constrained Parameters}\label{sec:constgauge}

Now we will take into account the presence of irreducible gauge symmetries of the system. Let us
consider the case when the gauge symmetry parameters are not arbitrary, but obey some differential
constraints. In other words, we have a set of gauge parameters $\ve^{\,\a}$,
with $\a=1,2,...,r$, governed by the following order-$\tilde{p}$ system of PDE's:
\beq T^a[\ve^{\,\a},\ve^{\,\a}_\m,\ldots,\ve^{\,\a}_{\m(\tilde{p})}]=0,\quad\text{with}
\quad a=1,2,...,\tilde{t}.\eeq{A20}
We further assume that the system~(\ref{A20}) is \underline{involutive}, and that the gauge symmetries
appear in a \underline{single finite order} $s$. The $k$-th derivatives of the gauge parameters evaluated
at $x=x_0$ constitute a set of monomials $\bar\ve_{\m(k)}^{\,\a}$. Because the gauge symmetries are
irreducible, the number of \underline{undetermined} monomials at $\mathcal{O}(k+s)$ follows directly
from Eq.~(\ref{A14}):
\beq \hat{n}^R_k~=~r{k+s+D-1\choose k+s}-\sum_n\left(\tilde t_n-\sum_j(-)^j\tilde l_{n,\,j}\right)
{k+s-n+D-1\choose k+s-n},\eeq{A21}
for large $k$, where $\tilde{t}_n$ is the number of equations at order $n$, and $\tilde{l}_{n,\,j}$
number of gauge identities at total order $n$ and reducibility order $j$. This count generalizes
Eq.~(\ref{A11}) to the case when the gauge parameters are governed by an involutive system of equations.

In order to find the number of $\mathcal{O}(k)$ monomials $\bar{\F}^A_{\m(k)}$ that are undetermined as
well as gauge inequivalent, we must subtract the count~(\ref{A21}) from the gauge-redundant count~(\ref{A14}).
To simplify the exercise we first note that the expansion~(\ref{A15}) gives:
\beq \hat{n}^R_k~=~{k+D-1\choose k}\left\{\tilde{\D}+\frac{2}{k}\,(D-1)\left(\tilde{\mathfrak{D}}
+\tfrac{1}{2}s\tilde{\mc}\right)+\mathcal{O}\left(\frac{1}{k^2}\right)\right\},\qquad k\rightarrow
\infty,\eeq{A22}
where $\tilde{\mc}$ and $\tilde{\mathfrak{D}}$ are respectively the compatibility coefficient and the
DoF count of the involutive system~(\ref{A20}) of the gauge parameters; they are given by:
\beq \tilde{\mc}=r-\sum_n\left(\tilde{t}_n-\sum_j(-)^j\tilde{l}_{n,\,j}\right),\qquad
\tilde{\mathfrak{D}}=\tfrac{1}{2}\sum_nn\left(\tilde{t}_n-\sum_j(-)^j\tilde{l}_{n,\,j}\right).\eeq{A23}
While $\tilde{\mc}=0$ by the assumption of absolute compatibility, $\tilde{\mathfrak{D}}$ counts the
number of \underline{pure gauge DoF} of the original system~(\ref{eq:eom}) that enjoys the
local gauge symmetry under consideration. A straightforward calculation now leads to the physical
DoF count:
\beq \mathfrak{D}=\tfrac{1}{2}\sum_{n}n\left(t_n-\sum_j(-)^jl_{n,\,j}\right)-\tilde{\mathfrak{D}}.
\eeq{A24}
This is an intuitively-clear generalization of Eq.~(\ref{A18}): the physical DoF count is obtained
simply by subtracting the pure-gauge DoF count from the dynamical DoF count (including gauge modes).
When gauge symmetries are absent, $\tilde{\mathfrak{D}}=0$, and we recover Eq.~(\ref{A18}).

\section{Involutive Deformations}\label{sec:invdeform}

Given a set of free field equations in the involutive form$-$with all the gauge identities and
symmetries identified$-$it is possible to systematically deform the theory and thereby introduce
consistent of interactions~\cite{KaLySh3}. The algebraic consistency and the correct DoF count
are obtained, even for the deformed system, by strictly preserving the involutive structure.
The same approach can be taken also for the problem of writing down consistent EoM's for fields
propagating freely in nontrivial backgrounds~\cite{Cortese:2013lda,Kulaxizi:2014yxa,Cortese:2017ieu}.
To see how this works, let us first enumerate the consistency conditions to be taken into account:
\begin{enumerate}
\item {\bf Algebraic Consistency:} The dynamical equations and constraints/gauge-fixing conditions ought
to be mutually compatible. They should not give rise to any new conditions on the fields that cease
to exist when the background is switched off~\cite{FP}.
\item {\bf Gauge Invariance:} When placed in a nontrivial background, the gauge symmetries of a dynamical
system should be preserved in order to eliminate unphysical modes.
\item {\bf No Higher Derivatives:} Constraint equations must not contain more than one time-derivatives
of the field, i.e., they cannot be promoted to dynamical ones. On the other hand, dynamical equations ought
to include two time-derivatives at most. Otherwise, the system will generically be plagued with Ostrogradsky
instability~\cite{Ostrogradksi} (see also~\cite{deRham:2016wji} for a recent discussion).
\item {\bf Hyperbolicity:} Even when the dynamical equations contain only up to two time-derivatives,
non-canonical kinetic terms may ruin the hyperbolicity of the system. In other words, such terms may render
the Cauchy problem ill posed~\cite{vz}.
\item {\bf Causality:}  A hyperbolic system of PDE's describing the dynamics of some field should also have
a propagation speed not exceeding the speed of light. When non-canonical kinetic terms are present in the
dynamical equations of a Lorentz-invariant theory, this feature cannot be taken for granted
(see~\cite{Rahman:2015pzl} for a recent review).
\item {\bf DoF Count:} Last but not the least, the count of physical DoF's of a dynamical system should be
correct. In other words, consistent free propagation in a nontrivial background implies that the DoF count
does not alter by turning off the background.
\end{enumerate}

In the involutive deformation method conditions $1,2$ and $6$ are automatically taken care of by the involutive
structure. By virtue of working at the EoM level, one also has conditions $3,4$ and $5$ under control, since
higher-derivatives and/or non-canonical kinetic terms can simply be avoided by choice. Lagrangian formulation
has severe limitations in this regard, as we already mentioned in the Introduction.

Below we outline the systematic procedure of writing down consistent EoM's for free higher-spin fields in
nontrivial gravitational or electromagnetic backgrounds.

\begin{itemize}
\item The flat-space free system of equations is written down in an involutive form.

\item All the gauge identities and gauge symmetries of the system are identified.

\item Zeroth-order deformation of the system, in the presence of a nontrivial background, is obtained by replacing
ordinary derivatives by covariant ones (minimal coupling).

\item Because covariant derivatives do not commute, zeroth-order deformations will not be self sufficient in general.
Higher-order deformations of the equations, gauge identities/symmetries will cast Eqs.~(\ref{eq:eom})--(\ref{gaugesymm})
into the following schematic form:
\bea T^a&=&T_0^a+gT_1^a+g^2T_2^a+\cdots,\nonumber\\
L^i_a&=&L^i_{a,\,0}+gL^i_{a,\,1}+g^2L^i_{a,\,2}+\cdots,\label{def2}\\
R^A_\a&=&R^A_{\a,\,0}+gR^A_{\a,\,1}+g^2R^A_{\a,\,2}+\cdots,\nonumber\eea{defff2}
where the numerical subscript denotes the deformation order in some \underline{dimensionless} parameter $g$.
In fact, the deformation parameter $g$ is just a book-keeping device to track the power of background curvature.
For example, linear terms in the curvature will be $\mathcal{O}(g)$, quadratic-curvature terms will be
$\mathcal{O}(g^2)$, and so on.

\item The deformations~(\ref{def2}) are chosen in such a way that the gauge identities and gauge symmetries
hold good order by order in $g$, and that the number equations and gauge identities/symmetries at a given
derivative order do no change\footnote{In principle, the derivative orders of the equations and gauge
identities/symmetries may increase at any order in $g$. We, however, do not explore this possibility in
order to make sure that the consistency conditions involving higher derivatives, hyperbolicity and causality
($3,4$ and $5$) are not violated.}.

\item Because derivatives and curvatures are dimensionful quantities, their higher powers must come with
suppression by a relevant mass scale $\L$. Accordingly, the respective mass dimensions of the
deformations~(\ref{def2}) remain the same at any order. In order for an effective field theory description
to make sense, $\L$ should be parametrically larger than any other mass scale in the system.
\end{itemize}

This method ensures that the system remains involutive and absolutely compatible, and contains the
same number of physical DoF's before and after the deformation. While algebraic consistency of the system
is guaranteed by the involutive structure, causal propagation is maintained by avoiding non-canonical kinetic
terms in the dynamical equations.

\section{Technical Details}\label{sec:technical}

Here we provide some technical details omitted in the bulk of the article for the sake of readability. Appendix~\ref{sec:tech1} deals with gravitational backgrounds, whereas~\ref{sec:tech2} with EM backgrounds. They present
some useful formulae and elaborate on important technical steps leading to some of the derivations for both bosonic and fermionic fields.

\subsection{Gravitational Background}\label{sec:tech1}

The Riemann tensor can be decomposed into the following irreducible pieces:
\beq R_{\m\n\r\s}=W_{\m\n\r\s}+\left(\tfrac{2}{D-2}\right)\left(g_{\m[\r}S_{\s]\n}-g_{\n[\r}S_{\s]\m}\right)
+\tfrac{2}{D(D-1)}Rg_{\m[\r}g_{\s]\n},\eeq{C1}
where $D$ is the space-time dimensionality.
Note that a conformally flat Einstein manifold is a maximally symmetric space. For a maximally symmetric space, one can write:
\beq R_{\m\n\r\s}=-\tfrac{1}{L^2}\left(g_{\m\r}g_{\n\s}-g_{\m\s}g_{\n\r}\right),\qquad
R_{\m\n}=-\left(\tfrac{D-1}{L^2}\right)g_{\m\n},\qquad R=-\tfrac{D(D-1)}{L^2},\eeq{C2}
where $L$ is the AdS radius (for dS space, we make the substitution: $L^2\rightarrow -L^2$). Then, the commutator of covariant
derivatives~(\ref{1.4})--(\ref{1.5}) reduces to the following form:
\beq [\nb_\m,\nb_\n]=\left\{
                          \begin{array}{ll}
                                 -\rad\left(2u_{[\m}d_{\n]}\right), & \hbox{for bosons;} \\
                                 -\rad\left(2u_{[\m}d_{\n]}+\tfrac{1}{2}\g_{\m\n}\right), & \hbox{for fermions.}
                          \end{array}
                    \right.\eeq{C3}
The commutator of divergence and symmetrized gradient in this case reads:
\beq [\ddiv,\dgrad]=\left\{
                          \begin{array}{ll}
                                 \nb^2-\rad N\left(N+D-2\right)+\rad u^2d^2, & \hbox{for bosons;} \\
                                 \nb^2-\rad N\left(N+D-\tfrac{3}{2}\right)+\rad\left(u^2d^2+\tfrac{1}{2}\us\,\ds\right), & \hbox{for fermions.}
                          \end{array}
                    \right.\eeq{C4}

\subsubsection*{Computations with Bosonic Fields}

The derivation of the explicit form of Eq.~(\ref{master}) relies on the following
commutators:
\bea
&{[}\ddiv,\nb^2{]}=-2R_{\m\n\r\s}\nb^\m u^\r d^\n d^\s+R_{\m\n}\nb^\m d^\n
+(\nb^\m R_{\m\n})d^\n-2\nb_{[\m} R_{\n]\r}u^\m d^\n d^\r,&\nonumber\\
&{[}\ddiv, R_{\m\n}u^\m d^\n{]}=(\nb_\r R_{\m\n})u^\m d^\n d^\r
+(\nb^\m R_{\m\n})d^\n+R_{\m\n}\nb^\m d^\n,&\label{C5}\\
&{[}\ddiv,R_{\m\n\r\s}u^\m u^\r d^\n d^\s{]}=(\nb_\a R_{\m\n\r\s})u^\m u^\r d^\a d^\n d^\s + 4\nb_{[\m} R_{\n]\r}u^\m d^\n d^\r +2R_{\m\n\r\s}\nb^\m u^\r d^\n d^\s.&\nonumber\eea{C55555}
With the help of these commutators, it is easy to obtain the following:
\begin{equation}\label{C6}\begin{split}
[\hat{g}_1,\hat{g}_0]&=2(\a_1-1)R_{\m\n\r\s}\nb^\m u^\r d^\n d^\s+(\a_2+1)R_{\m\n}\nb^\m d^\n-R_{\m\n\r\s}u^\m u^\r d^\n
d^\s\,[\a_1,\ddiv]\\&-R_{\m\n}u^\m\,[\a_2,\ddiv]-R\,[\a_3,\ddiv]
+[M^2,\ddiv]+\a_1(\nb_\a R_{\m\n\r\s})u^\m u^\r d^\a d^\n d^\s\\
&+2(2\a_1-1)(\nb_{[\m} R_{\n]\r})u^\m d^\n d^\r+\a_2(\nb_\m R_{\n\r})
u^\r d^\m d^\n+\a_3(\nb_\m R)d^\m+\otwo.\end{split}\end{equation}
Eq.~(\ref{master}) then follows from the decomposition~(\ref{C1}).
Terms containing gradients of the curvature can be further massaged with the
decomposition given in Eqs.~(\ref{tableaux})--(\ref{XYZUdefined}).

In order to prove Eq.~(\ref{2.19}), let us note the combination $\left(R_{\m\n\r\s}u^\m u^\r d^\n d^\s-R_{\m\n}u^\m d^\n\right)$ commutes with
the trace operator, which is easy to show. Then, with the choices~(\ref{2.140})
the commutator $[\hat{g}_2,\hat{g}_0]$ reduces to the following:
\beq [\hat{g}_2,\hat{g}_0]=-R[\a_3,d^2]+[M^2,d^2]+\otwo,\eeq{C7}
which gives rise the relation~(\ref{2.19}) for the choices and~(\ref{2.15})--(\ref{2.16}).

In (\ref{2.23}) we used the operators $\hat{\mathcal{O}}_i, i=1,2,3$, without spelling out their explicit forms; these operators are defined as follows:
\bea
&\hat{\mathcal{O}}_3=[\hat{g}_1,\hat{g}_2],&\nonumber\\
&\hat{\mathcal{O}}_2=[\hat{g}_2,\hat{g}_0]+\tfrac{2}{(D-1)(D+2)}\,\hat{g}_2 R\left(2N+D-2\right)+\hat{g}_2P(N),&\label{C8}\\
&\hat{\mathcal{O}}_1=[\hat{g}_0,\hat{g}_1]-\tfrac{2}{(D-1)(D+2)}\left[2\hat{g}_1R\left(2N+D-1\right)
-\hat{g}_2\left(u^2d\!\cdot\!U+u\!\cdot\!U\right)\right]-\hat{g}_1Q(N).&\nonumber
\eea{C88888}

Next, we move on to the massless case and give the explicit expression of the weight-$0$ operator
$\mathcal{X}_0$ appearing in Eq.~(\ref{4.22}); it reads:
\beq \mathcal{X}_0=-2\left(R_{\m\n\r\s}u^\m u^\r d^\n d^\s-R_{\m\n}u^\m d^\n\right)
-\tfrac{2(N-1)(N+D-2)}{(D-1)(D+2)}\,R+M_0^{\prime\,2}.\eeq{C9}

Then, we consider the details of partially-massless bosons in Section~\ref{sec:PMb}. To avoid clumsiness in
the expressions, in what follows we will set the AdS radius to unity: $L=1$. The following
commutation relations involving the d'Alembertian operator are useful:
\bea
&\left[\Box\!-\!aN\!-\!b, \dgrad\right]=-\dgrad\left\{a+2(2N+D-1)\right\}
+4u^2\ddiv,&\label{C10}\\
&\left[\Box\!-\!aN\!-\!b, (\dgrad)^2\!-\!cu^2\right]=-2(\dgrad)^2(a\!+\!4N\!+\!2D)
\!+\!8u^2\dgrad\ddiv\!+\!4u^2(\Box\!+\!\tfrac{1}{2}ac),&~~~~~~\eea{C11}
where $a,b$ and $c$ are numerical constants. For the divergence operator, note from Eq.~(\ref{6.3}) that,
by definition: $[\ddiv,\dgrad]=\Box$. We also have the following important commutator:
\beq \left[\ddiv,\,(\dgrad)^2-cu^2\right]\,=\,2\dgrad \left\{\Box-c-(2N+D-1)\right\}+4u^2\ddiv.\eeq{C12}
Last but not the least, the trace operator has the commutation relations:
\beq \left[d^2,\dgrad\right]\,=\,2\ddiv,\qquad \left[d^2,\,(\dgrad)^2-cu^2\right]\,=\,4\dgrad\,\ddiv
+2\left\{\Box-c(2N+D)\right\}.\eeq{C13}
The variations of the left-hand sides of EoM's for the case $k=1$ are given by:
\begin{equation}\label{C14}\begin{split}
\d\!\left[\hat{g}_0^{(2)}\F_s^{(2)}\right]=\left\{(\dgrad)^2\mathcal{L}_{11}
+u^2\mathcal{Q}_{11}\right\}\l_{s-2},\qquad\\
\d\!\left[\hat{g}_1\F_s^{(2)}\right]=\left\{\dgrad\mathcal{L}_{12}\right\}\l_{s-2},
\qquad \d\!\left[\hat{g}_2\F_s^{(2)}\right]=\left\{\mathcal{L}_{13}\right\}\l_{s-2},
\end{split}\end{equation}
where we recall that $L=1$, and the $\mathcal{L}$'s and $\mathcal{Q}$'s are the following linear functions of $N$:
\begin{equation}\label{C15}\begin{split}
\mathcal{L}_{11}&=(a'_1-a_1-8)N+[b_1'-b_1-2(a_1+2D)],\\
\mathcal{Q}_{11}&=[4(a'_1-2)-c_1(a_1'-a_1-8)]N+[4b_1'-c_1(b_1'-2a_1-b_1)],\\
\mathcal{L}_{12}&=2(a'_1-2)N+2(b_1'-c_1-D+1),\\
\mathcal{L}_{13}&=2(a'_1-2c_1)N+2(b_1'-Dc_1).
\end{split}\end{equation}
Similarly, the variations for a depth-$3$ PM field, corresponding to $k=2$, read:
\bea
&\d\!\left[\hat{g}_0^{(3)}\F_s^{(3)}\right]=\left\{(\dgrad)^3\mathcal{L}_{21}
+u^2\dgrad\mathcal{Q}_{21}\right\}\l_{s-3},&\nonumber\\
&\d\!\left[\hat{g}_1\F_s^{(3)}\right]=\left\{(\dgrad)^2\mathcal{L}_{22}
+u^2\mathcal{Q}_{22}\right\}\l_{s-3},&\label{C16}\\
&\d\!\left[\hat{g}_2\F_s^{(3)}\right]=\left\{\dgrad\mathcal{L}_{23}\right\}\l_{s-3},&\nonumber
\eea{C166666}
where again the $\mathcal{L}_{2i}$'s and $\mathcal{Q}_{2i}$'s are linear functions of $N$, given by:
\begin{equation}\label{C17}\begin{split}
\mathcal{L}_{21}&=(a'_2-a_2-12)N+[b_2'-b_2-3(a_2+2(D+1))],\\
\mathcal{Q}_{21}&=[4(3a'_2-2c_2-4)-c_2(a'_2-a_2-12)]N\\
&~~~+[c_2(3a_2+b_2+2D-2)-b_2'(c_2-12)-8(D-1)],\\
\mathcal{L}_{22}&=3(a'_2-4)N+(3b_2'-2c_2-6D+2),\\
\mathcal{Q}_{22}&=-(c_2-4)(a'_2N+b_2'),\\
\mathcal{L}_{23}&=2(3a'_2-2c_2-4)N+[(3b_2'-(D+2)(c_2+2)+6].
\end{split}\end{equation}
Next, we elaborate on the computations with fermionic fields in gravitational backgrounds.

\subsubsection*{Computations with Fermionic Fields}

In deriving Eq.~(\ref{3.14}), one can first make use of the commutator~(\ref{1.5}) to write:
\beq [\nbs,\ddiv]=R_{\m\n\r\s}\g^\m u^\r d^\n d^\s-R_{\m\n}\g^\m d^\n
-\tfrac{1}{4}R_{\m\n\r\s}d^\m(\g^\n\g^{\r\s}).\eeq{C18}
Thanks to the $\g$-matrix identity: $\g^\n\g^{\r\s}=\g^{\n\r\s}+2\h^{\n[\r}\g^{\s]}$, and the properties of the Riemann tensor, the last term in the above equation simplifies to $\tfrac{1}{2}R_{\m\n}\g^\m d^\n$. Then, one can plug in the Riemann-tensor decomposition~(\ref{C1}) to arrive at Eq.~(\ref{3.14}).

Next, we give the technical details of PM fermions in Section~\ref{sec:PMf}.
Here, the AdS radius is set to unity: $L=1$. It is important to note that, unlike the usual covariant derivative $\nb_\m$,
the deformed one $\D_\m$ does not commute with $\g$-matrices. To be explicit:
\beq [\g^\m,\D^\n]=[\D^\m,\g^\n]=-\g^{\m\n}.\eeq{C20.00}
Some commutators involving the Dirac operator that will be useful for our purpose are:
\bea
\left[\Ds-aN-b,\Dgrad\right]&=&-(a+1)\Dgrad+\us\left(\Ds-N+\us\,\ds\right),\label{C19}\\
\left[\Ds-aN-b,(\Dgrad)^2-cu^2\right]&=&-2(a+1)(\Dgrad)^2+2u^2\left(\Ds-N+ac+\us\,\ds\right)
\nonumber\\&&+2\us\,\Dgrad\left(\Ds-N-1+\us\,\ds\right),\eea{C20}
where $a,b$ and $c$ are numerical constants. Similarly, for the divergence operator:
\bea
\left[\Ddiv,\Dgrad\right]&=&\left(\Ds+N+D-1\right)\left(\Ds-N\right)
+\left(\us+u^2\ds\right)\ds,\label{C21}\\
\left[\Ddiv,(\Dgrad)^2-cu^2\right]&=&2\Dgrad\left(\Ds+N+D\right)\left(\Ds-N-1\right)-2(c-1)\Dgrad\nonumber\\
&&+4u^2\Ddiv+2\Dgrad\left(\us+u^2\ds\right)\ds.\eea{C22}
The gamma-trace operator, on the other hand, has the commutation relations:
\bea \left[\ds,\Dgrad\right]&=&\left(\Ds-N\right)+\us\,\ds,\label{C23}\\ \left[\ds,(\Dgrad)^2-cu^2\right]&=&2\left(\Dgrad+\us\right)
\left(\Ds-N-1+\us\,\ds\right)-2(c-1)\us.\eea{C24}

First, we compute the variations of the left-hand sides of EoM's for a depth-2 PM fermion, which corresponds to $k=1$.
Given the commutation relations~(\ref{C20}),~(\ref{C22}) and~(\ref{C24}), and the involutive system
of the gauge parameter, they reduce to:
\bea
&\d\!\left[\hat{f}_0^{(2)}\Ps_n^{(2)}\right]=\left\{(\Dgrad)^2\,\mathcal{P}_{11}
+\us\,\Dgrad\,\mathcal{M}_{11}+u^2\mathcal{N}_{11}\right\}\ve_{n-2},&\nonumber\\
&\d\!\left[\hat{g}_1^{\,\prime}\Ps_n^{(2)}\right]=\left\{\Dgrad\,\mathcal{P}_{12}
+\us\,\mathcal{M}_{12}\right\}\ve_{n-2},&\label{C25}\\
&\d\!\left[\hat{f}_1\Ps_n^{(2)}\right]=\left\{\Dgrad\,\mathcal{P}_{13}+\us\,
\mathcal{M}_{13}\right\}\ve_{n-2},\nonumber\eea{C2555555}
where we set $L=1$, and the $\mathcal{P},\,\mathcal{M}$ and $\mathcal{N}$'s are the following polynomial functions of $N$:
\bea
&\mathcal{P}_{11}~=~(\a'_1-\a_1)N+[\,\b_1'-\b_1-2(\a_1+1)\,],&\nonumber\\
&\mathcal{P}_{12}~=~2(\a^{\prime\, 2}_1\!-\!1)N^2+2[\a'_1(2\b'_1\!+\!D\!-\!1)\!-\!D\!-\!1]N
+[2(\b'_1\!+\!D)(\b'_1\!-\!1)\!-\!2(\d_1\!-\!1)],&\nonumber\\
&\mathcal{P}_{13}~=~\mathcal{M}_{11}~=~2(\a'_1-1)N+2(\b_1'-1),&\label{C26}\\
&\mathcal{M}_{12}~=~0,\qquad\mathcal{M}_{13}~=~2(\a'_1-1)N+[2(\b_1'-1)-2(\d_1-1)],&\nonumber\\
&\mathcal{N}_{11}~=~[2(\a'_1-1)-(\a'_1-\a_1)\d_1]N+[2(\a'_1+1)-(\b'_1-\b_1)\d_1].&
\nonumber\eea{C2666666}
Similarly, the variations for a depth-$3$ PM field, corresponding to $k=2$,
are given by:
\bea
&\d\!\left[\hat{f}_0^{(2)}\Ps_n^{(3)}\right]=\left\{(\Dgrad)^3\,\mathcal{P}_{21}
+\us(\Dgrad)^2\mathcal{M}_{21}+u^2\Dgrad\,\mathcal{N}_{21}+u^2\us\,
\mathcal{R}_{21}\right\}\ve_{n-3},&\nonumber\\
&\d\!\left[\hat{g}_1^{\,\prime}\Ps_n^{(3)}\right]=\left\{(\Dgrad)^2\mathcal{P}_{22}
+\us\,\Dgrad\,\mathcal{M}_{22}+u^2\mathcal{N}_{22}\right\}\ve_{n-3},&\label{C27}\\
&\d\!\left[\hat{f}_1\Ps_n^{(3)}\right]=\left\{(\Dgrad)^2\mathcal{P}_{23}
+\us\,\Dgrad\,\mathcal{M}_{23}+u^2\mathcal{N}_{23}\right\}\ve_{n-3},&\nonumber
\eea{C277777}
where again the $\mathcal{P},\,\mathcal{M},\,\mathcal{N}$ and $\mathcal{R}$'s  are polynomial functions of $N$,
given by:
\bea
&\mathcal{P}_{21}~=~(\a'_2-\a_2)N+[\,\b_2'-\b_2-3(\a_2+1)\,],&\nonumber\\
&\mathcal{P}_{22}~=~3(\a^{\prime\, 2}_2\!-\!1)N^2+3[\a'_2(2\b'_2\!+\!D\!-\!1)\!-\!D\!-\!3]N
+[3(\b'_2\!+\!D\!+\!1)(\b'_2\!-\!2)\!-2(\d_2\!-\!4)],&\nonumber\\
&\mathcal{P}_{23}~=~\mathcal{M}_{21}~=~3(\a'_2-1)N+3(\b_2'-2),&\nonumber\\
&\mathcal{M}_{22}~=~0,\qquad\mathcal{M}_{23}~=~6(\a'_2-1)N+[6(\b_2'-2)-2(\d_2-4)],&\label{C28}\\
&\mathcal{N}_{21}~=~[6(\a'_2-1)-(\a'_2-\a_2)\d_2]N+[6\b'_2-4-(\b'_2-\b_2-3\a_1-1)\d_2],&
\nonumber\\
&\mathcal{N}_{22}~=~(4-\d_2)[(\a'_2(2\b'_2+D-1)-D+1)N+\b'_2(\b'_2+D-1)],&
\nonumber\\
&\mathcal{N}_{23}~=~\mathcal{R}_{21}~=~(4-\d_2)[(\a'_2-1)N+\b'_2].&
\nonumber\eea{C288888}
This finishes our exposition of the computational details for gravitational backgrounds.

\subsection{Electromagnetic Background}\label{sec:tech2}

Let us emphasize that minimal coupling to the EM background has been assumed. Here, the commutator of covariant derivatives
acts the same way on bosons and fermions:
\beq [\mD_\m,\mD_\n]\F=iqF_{\m\n}\F,\qquad [\mD_\m,\mD_\n]\Ps=iqF_{\m\n}\Ps.\eeq{C30}
Below we elaborate on some computations involving bosonic and fermionic fields.

\subsubsection*{Computations with Bosonic Fields}

The derivation Eq.~(\ref{2.36}) makes use of the following commutation relation:
\beq [d\!\cdot\!\mD,\mD^2]=-2iqF_{\m\n}\mD^\m d^\n-iqd_\m V^\m,\eeq{C31}
which simplifies the commutator $[\bar{g}_1,\bar{g}_0]$ to the following form:
\begin{equation}\label{C32}\begin{split}
[\bar{g}_1,\bar{g}_0]&=iq\left(\a-2\right)F_{\m\n}\mD^\m d^\n-iq\a \de_{(\m}F_{\n)\r}u^\r d^\m d^\n+iq(\a-1)d\!\cdot\!V
\\&~~~-iq[\a,d\!\cdot\!\mD]F_{\m\n}u^\m d^\n+[\bar{M}^2,d\!\cdot\!\mD]+\otwob.
\end{split}\end{equation}
Then, one can easily arrive at Eq.~(\ref{2.36}) from the definition of $A_{\m\n\r}$ given in Eq.~(\ref{AVdefined}).

\subsubsection*{Computations with Fermionic Fields}

We will now provide justification for the ans\"atze~(\ref{3.32})--(\ref{3.33}). At first order in $F_{\m\n}$, the non-minimal
deformation $\mathcal{A}$ of the Dirac operator may contain five independent terms:
\beq \mathcal{A}=iq\left(a_{+}F^{+}_{\m\n}+a_{-}F^{-}_{\m\n}+a_1F_{\m\r}\g^\r\g_\n+a_2F_{\n\r}\g^\r\g_\m\right)u^\m d^\n
+iqa_0 F_{\r\s}\g^{\r\s}+\cdots,\eeq{C33}
where the $a$'s are weight-$0$ operators of mass dimension $-1$, and the ellipses stand for terms containing derivatives
or higher powers of the field strength. The third term on the right-hand side of Eq.~(\ref{C33}) is however
redundant since it is proportional to $\bar{f}_1$, under the assumption~(\ref{3.31}). Without any loss of generality
therefore one can set: $a_1=0$. Given this, if one further requires that the Dirac operator be hermitian in the sense
of footnote~5, one must also set: $a_2=0$. This justifies our ansatz~(\ref{3.32}).
Similarly, the non-minimal deformation $\mathcal{B}$ of the divergence operator takes the generic form:
\beq \mathcal{B}=iq\left(b_0 F_{\m\n}\g^\m d^\n+b_1F_{\r\s}\g^{\r\s}\ds\right)+\cdots,\eeq{C34}
with $b_0$ and $b_1$ being weight-$0$ operators of dimension $-1$, and the ellipses contain derivatives and higher powers
of the field strength. Again, without any loss if generality, one can set: $b_1=0$. This leads us to the ansatz~(\ref{3.33}).

Next, we compute the graded commutators of Section~\ref{sec:mfem}, which are eventually expressed in Eq.~(\ref{3.36}).
Starting from Eq.~(\ref{3.28}), a straightforward computation gives:
\begin{equation}\label{com1}\begin{split}
[\bar{f}_0,\bar{g}_1]&=iq\left[1-m(a_{+}-a_{-}+2b_0)\right]F_{\m\n}\g^\m d^\n+2iq\left(a_{-}-b_0\right)F_{\m\n}d^\m\mD^\n
\\&~~~+iq\left(a_{+}-a_{-}\right)\left[F_{\m\n}\g^\m\mD^\n\bar{f}_1+\tfrac{1}{2}F_{\r\s}\g^{\r\s}\left(\bar{g}_1
-\slashed{\mD}\bar{f}_1\right)\right]\\&~~~-iq\left(a_{+}-a_{-}+2b_0\right)F_{\m\n}\g^\m d^\n\bar{f}_0+\cdots,
\end{split}\end{equation}
where the ellipses stand for terms containing derivatives or higher powers of the field strength, and commutators involving
the weight-$0$ operators $a_{\pm}$, $a_0$ and $b_0$. In deriving the above result, we have used a number of $\g$-matrix
identities, in particular:
\beq F^{+\m\n}=\tfrac{1}{4}\left(\g^\m\g^{\r\s}\g^\n-\g^\n\g^{\r\s}\g^\m\right)F_{\r\s},\qquad
F^{-\m\n}=-\tfrac{1}{4}\left(\g^{\m\n}\g^{\r\s}+\g^{\r\s}\g^{\m\n}\right)F_{\r\s}.\eeq{C36}
On the other hand, Eq.~(\ref{3.29}) leads rather easily to the following result:
\beq [\bar{g}_1,\bar{f}_1]=2b_0F_{\m\n}\g^\m d^\n\bar{f}_1+\cdots.\eeq{com2}
Finally, in order to work out $\{\bar{f}_1,\bar{f}_0\}$ from Eq.~(\ref{3.30}), we need to compute the anti-commutator
$\{\ds,\mathcal{A}\}$ with the help of the following $\g$-matrix identities:
\begin{equation}\label{C38}\begin{split}
\g^{\m\n\r\s}\g^\l+\g^\l\g^{\m\n\r\s}=2\g^{\m\n\r\s\l},\qquad \g_\m\g^{\m\n\r\s}=(D-3)\g^{\n\r\s},\\
\g^{\m\n}\g^\r+\g^\r\g^{\m\n}=2\g^{\m\n\r},\qquad \g^{\m\n\r}=\g^{\m\n}\g^\r+2\h^{\r[\m}\g^{\n]}.~~~
\end{split}\end{equation}
After a straightforward calculation, one arrives at the following expression:
\begin{equation}\label{ancom}\begin{split}
\{\bar{f}_1,\bar{f}_0\}&=2\bar{g}_1-2m\bar{f}_1-iq\left[(D-4)a_{+}-(D-2)a_{-}+4a_0+2b_0\right]F_{\m\n}\g^\m d^\n
\\&~~~+iqF_{\m\n}\left[2(a_{+}+a_{-})u^\n d^\n+\tfrac{1}{2}\{4a_0+(D-3)(a_{+}-a_{-})\}\g^{\m\n}\right]\bar{f}_1+\cdots.
\end{split}\end{equation}

Clearly, the $F_{\m\n}\g^\m d^\n$- and $F_{\m\n}\g^\m\mD^\n$-terms appearing in the first lines of Eqs.~(\ref{com1})
and~(\ref{ancom}) obstruct the closure of these commutators, for spin $s\geq\tfrac{3}{2}$. Their coefficients must
therefore be set to zero, which results in the choice~(\ref{3.34}). At $\ooneq$, other offending terms may
appear through derivatives of the field strength. Omitted in the ellipses of
Eqs.~(\ref{com1}),~(\ref{com2}) and~(\ref{ancom}), such terms can be eliminated by the condition~(\ref{3.35}).

We finish with the derivation of Eq.~(\ref{3.39}). Because the non-minimal corrections to the gauge-identity
operators~(\ref{3.37}) are proportional to the EoM's, it is easy to see why the schematic form~(\ref{3.39}) should
appear. After a somewhat tedious computation, one finds that the operators $\bar{O}_2$, $\bar{O}_1'$ and $\bar{O}_1$
are given by:
\bea &\bar{O}_2=\bar{h}_2-\left(i\e q/m\right)F_{\m\n}\g^\m d^\n\bar{f}_1,&\nonumber\\
&\bar{O}_1'=\bar{h}_1'+\left(2iq/m\right)F_{\m\n}\left(\e\g^\m d^\n+u^\m d^\n\bar{f}_1\right),&\label{C40}\\
&\bar{O}_1=\bar{h}_1\!+\!\left(2iq/m\right)\!F_{\m\n}\!\left[u^\m d^\n\bar{g}_1\!+\!\e\g^\m\mD^\n\bar{f}_1
\!+\!\tfrac{1}{2}\e\g^{\m\n}\bar{f}_1\slashed{\mD}\!+\!\left(1\!-\!\tfrac{3}{2}\e\right)\g^\m d^\n\bar{f}_0
\!-\!(2\!-\!\e)\slashed{\mD}\g^\m d^\n\right].&\nonumber\eea{C400000}
This marks the end of the necessary technical details.

\end{appendix}

\end{document}